# Multipath Mitigation Technology-integrated GNSS Direct Position Estimation Plug-in Module

Sergio Vicenzo[1] (ORCID: 0000-0003-2974-7899), Bing Xu[1]

[1]Department of Aeronautical and Aviation Engineering, The Hong Kong Polytechnic University, Hong Kong SAR, China

*Correspondence: Bing Xu (E-mail: pbing.xu@polyu.edu.hk), The Hong Kong Polytechnic University, Hong Kong SAR, China

## Abstract

Non-line-of-sight (NLOS) and multipath (MP) signals have long been the two major issues for accurate Global navigation satellite system (GNSS) positioning in urban environments. Direct position estimation (DPE) is an effective solution to the MP issue at the signal processing level. Unlike two-step positioning (2SP) receivers, DPE directly solves for the receiver position, velocity, and time (PVT) in the navigation domain, without the estimation of intermediate measurements, and thus allowing it to provide more robust and accurate PVT estimates in the presence of MP and weak signals. Even though the advantages of DPE have been established, GNSS positioning with DPE is mostly left unapplied commercially, and continuing research into DPE has remained relatively stagnant over the past few years. To encourage further research on DPE by the GNSS community, we propose a DPE plug-in module that can be integrated into the conventional 2SP software-defined receivers (SDRs). Programmed in a user-friendly language, MATLAB, the proposed DPE plug-in module is aimed for better understanding and familiarity of a practical implementation of DPE. Its plug-in module architecture allows it to be incorporated with 2SP MATLAB SDRs, both vector tracking and scalar tracking with minimum changes, making it easy to use, and provides greater flexibility for researchers using various 2SP SDRs. The proposed DPE module has been evaluated with both real and simulated GNSS data replicating urban environments and has been compared against 2SP positioning. Results point out that the proposed DPE methodology manages to consistently provide superior performance against 2SP, especially under NLOS conditions. We propose to further improve the performance of DPE against MP by incorporating Multipath Mitigation Technology (MMT) into DPE. Referred to as MMT-DPE, it is proposed as a variant of DPE that adds the MP components into DPE signal model by integrating MMT cost function into DPE, with the aim to better suit DPE for urban environment applications. Results show that while in MP-only conditions, an MMT-integrated 2SP (MMT-2SP) has similar performance with MMT-DPE, the proposed MMT-DPE manages to show great superiority against NLOS, making it the preferable option for applications in urban environments.

## Introduction

The Global Navigation Satellite System (GNSS) is among the most highly relied on navigation and positioning systems in the world. From opening Google Maps to explore new cities during vacations to driving autonomous cars, most layman people do not realize how much they rely on GNSS in their everyday lives. That being said, GNSS often falters in densely populated areas, such as Hong Kong and Tokyo. High-rise buildings and other urban structures typically reflect GNSS signals before reaching the receiver, generating errors in the pseudorange measurements that may induce errors up to 50 meters (Hsu 2018). When only the reflected signal is received, it is classified as a non-line-of-sight (NLOS) reception. If a combination of the line-of-sight (LOS) and NLOS is received, it is categorized as a multipath (MP) reception. NLOS occurs when the receiver does not have a LOS view to the satellite, while MP signals are usually received when the receiver has LOS view to the satellite but is located close to urban structures e.g., buildings.

Various research has delved into solving the issue of MP and NLOS in GNSS positioning. Several researchers have proposed using 3D mapping-aided (3DMA) GNSS to identify and correct MP and NLOS errors. For example, Ng et al. (2020) proposed the use of 3DMA to generate a skymask (a skymask-based 3DMA) and detect the reflection points for NLOS signals to correct the induced NLOS errors. Results pointed out that a skymask-based 3DMA manages to maintain a positioning accuracy within 10 metres of error in dense urban areas (Ng et al. 2020). A more recent study was to use single-differenced residuals to generate a range residual map to solve for MP/NLOS by analysing the graphical characteristics of MP/NLOS errors (Xu et al. 2024).

The use machine learning methods to detect and mitigate MP errors have also been previously investigated. Phan et al. (2012) employed a support vector regression (SVR) to estimate multipath error by making use of the function between multipath error and satellite relative elevation angle and azimuth angle. On the other hand, Orabi et al. (2020) proposed a neural network-based delay-locked loop (NNDLL), which builds a neural network discriminator using multilayer perceptron (MLP) to estimate code phase directly. Li et al. (2022) also proposed a Deep Neural Network (DNN)-based correlators to mitigate MP effects at two-step positioning (2SP) tracking level through “cleaning” the Autocorrelation function (ACF) from distortions caused by MP.

Closas et al. (2007), on the other hand, proposed to solve the problem at the signal processing level by introducing direct position estimation (DPE). Differing from the conventional 2SP which first estimates the intermediate parameters i.e., pseudoranges and pseudorange rate to produce a position, velocity, and timing (PVT) estimates, DPE estimates the PVT “direct”-ly in the navigation domain using the maximum likelihood principle. By non-coherently combining the satellites correlations, DPE has been shown to offer more accurate PVT estimates against 2SP in cases of weak signals and MP (Axelrad et al. 2009, 2011; Closas and Gusi-Amigó 2017; Closas et al. 2009, 2015). However, GNSS positioning with DPE is mostly left uninvestigated and unapplied commercially, with only one notable commercial receiver that uses DPE (Dampf et al. 2018). One of the possible reasons is its difficult practical implementation. To encourage further research on DPE by the GNSS community, an open-source DPE implementation is proposed. To the best of the author’s knowledge, there has been only one other open-sourced DPE receiver, that is from Peretic and Gao (2021), which was programmed in Python and C++.

Instead of proposing a standalone software-defined receiver (SDR), DPE was proposed as a plug-in module that can be easily integrated into open-sourced 2SP MATLAB SDRs. Tracking data from 2SP is used to propagate the channel for DPE. The aim of the proposed DPE plug-in module is for the familiarization and popularization of DPE in the GNSS community. Proposing DPE as a "plug-in module" permits ease of use for those unfamiliar with DPE, and greater flexibility, since it can be integrated into any MATLAB 2SP SDR, instead of being a separate SDR like Peretic and Gao's (2021) DPE. Even though the proposed plug-in module is programmed using MATLAB, users can easily translate the code to be integrated into other 2SP SDRs in other programing languages. The DPE plug-in module is currently integrated with the GPS L1 C/A SoftGNSS MATLAB 2SP Scalar Tracking Loop (STL) SDR by Borre et al. (2007), and has been made available at GitHub (https://github.com/Sergio-Vicenzo/GPSL1-DPEmodule). Even though the proposed DPE is currently integrated into an STL SDR, integration of the DPE module is not restricted to an STL-based 2SP SDR and it is completely possible to integrate the DPE module into SDRs with other 2SP algorithms such as vector tracking loop (VTL). In addition, it is extandable to other constellations such as BeiDou.

Though DPE has been proven to be robust against MP, previous research has proved that its superior performance against 2SP typically falters in deep urban environments where MP and NLOS are the majority of signals received (Vicenzo et al. 2023). This does not necessarily mean that the performance of DPE is worse than 2SP, but rather that its performance is depreciated to a large degree with increasing errors from MP and NLOS the same way 2SP does. Tang et al. (2024) had also recently showed that while DPE remains more accurate in comparison to 2SP in harsh cases such as 4 out of 8 satellites being MP, DPE error still reaches up to tens, or even hundreds of meters with NLOS measurements, which makes it definitely unsuitable for urban positioning. With research into mitigating MP with 2SP has been widely explored, the prevalence and need for DPE as a robust positioning method against MP has diminished significantly, especially with its high computational load and inapplicability to commercial receivers producing RINEX-level measurements.

To solve this issue for DPE, the second contribution is to introduce a Multipath Mitigation Technology (MMT)-integrated DPE, nicknamed MMT-DPE. MMT was introduced as an efficient maximum likelihood estimator for accurate estimation of the code delays and carrier phase of the LOS and reflected signal (Weill 2002). In 2SP, the natural way to apply MMT is to integrate it at the tracking stage, replacing the discriminator to estimate tracking error.

On the other hand, an MMT-DPE would incorporate the cost function of MMT into DPE. In essence, both DPE and MMT are the same maximum likelihood estimators that share similar cost functions, with the difference that in MMT, a reflected path is assumed to be received in addition to the LOS path. Replacing DPE cost function with that of MMT would virtually mean that the MP component is incorporated into DPE signal model. Since the proposed DPE implementation made use of 2SP tracking to propagate the channel, MMT is applied into 2SP tracking, in which the code delays as well as the complex amplitudes are estimated. These estimates would later be used by MMT-DPE in positioning to compensate the MP measurements for DPE cost function.

The performance of the proposed MMT-DPE is evaluated with real GNSS urban data, as well as simulated urban datasets produced by a high-end GNSS simulator. Its performance is also compared to an MMT-integrated scalar tracking Least

Squares (LS) positioning solution to represent an MMT-integrated 2SP (MMT-2SP). Similarly with the conventional DPE plug-in module, MMT-DPE has been made available at the same GitHub page as the conventional DPE plug-in module.

In the following, the basic principles of DPE are introduced first, followed by the architecture of the proposed DPE plug-in module. Next, the principle of MMT is introduced, and the proposed MMT-DPE is presented. The methodology of data collection and results of the proposed DPE and MMT-DPE with both real and simulated GNSS data are then presented and discussed, continued with the conclusion of the findings.

## Modelling of Direct Position Estimation

The idea of DPE solving the PVT directly in the navigation domain is based on the fact that the code phase and Doppler frequency of a GNSS signal can be made as functions of the user PVT. Typically in 2SP, pseudoranges and pseudorange rates are estimated based on tracking measurements of the received signal and from there, the PVT is produced. The concept of DPE reverses this notion. Instead of estimating the received signal parameters i.e., code phase and Doppler frequency (through tracking in 2SP), the PVT acts directly as the variable to be estimated. As a specific PVT corresponds to a specific code phase and Doppler frequency, its corresponding signal replica can be generated and correlated with the incoming signal. Its correlation value determines how likely a candidate PVT is being the receiver PVT. Thus, in a way, rather than estimating the received signal parameters, DPE directly finds the PVT with the code phase and Doppler frequency that "matches" the best with that of the received signal. Recently, Tang et al. (2023) have noted that DPE can also be implemented with carrier phase measurements for greater positioning accuracy. But only the code phase and Doppler frequency measurements are considered for DPE here i.e., the original DPE implementation. This requires the generation of potential PVTs (candidate PVTs), which in this research, is initialized with 2SP PVT estimates (Closas and Gao 2020).

For this research, DPE uses the grid-based method, by which of establishing a set of candidate PVTs and obtaining the correlations from each satellite before finally non-coherently summing them up to obtain the candidate position with the highest signal correlation, which is considered the PVT estimate of DPE. The process is illustrated by the following cost function (Closas and Gao 2020).

$$\hat{\boldsymbol{\gamma}} = \arg\min_{\boldsymbol{\gamma},\mathbf{a}} \sum_{i=1}^{\mathrm{M}} \left\| \mathbf{x} - \mathbf{c}_i(\boldsymbol{\gamma})\mathrm{a}^i \right\|^2 \tag{1}$$

or equivalently

$$\hat{\boldsymbol{\gamma}} = \arg\max_{\boldsymbol{\gamma}} \sum_{i=1}^{\mathrm{M}} \| \mathbf{x}^{\mathrm{H}} \mathbf{c}_i(\boldsymbol{\gamma}) \|^2 \tag{2}$$

where $\hat{\boldsymbol{\gamma}} = [p_{\mathrm{x}}, p_y, p_z, \delta t]^{\mathrm{T}}$ are DPE estimates, in which $\mathbf{p} = [p_{\mathrm{x}}, p_y, p_z]$ is the receiver position in Earth-Centered-Earth-Fixed (ECEF) coordinates and $\delta t$ is the receiver clock bias. $\mathrm{a}^i$ is the complex amplitude for satellite *i*. $\mathbf{x}^{\mathrm{H}}$ is the Hermitian transpose of the received signal vector, denoted by $\mathbf{x} = \begin{bmatrix} \mathbf{x}[0] \\ \vdots \\ \mathbf{x}[\Delta T - \frac{1}{f_s}] \end{bmatrix}$ and for an L1 C/A signal, each element of $\mathbf{x}$ is samples of the received baseband signal, presented as the following (Peretic 2019).

$\mathrm{x}(n) = \sum_{i=1}^{M} A^i \mathrm{s}^i\{(f_{C/A} + f_{\mathrm{code,t}}^i)n + \phi_{\mathrm{code,t}}^i\}\exp\{j2\pi(f_{\mathrm{L1}} + f_{\mathrm{carr,t}}^i)n + \varphi_{\mathrm{carr,t}}^i\} + noise$ (3)

in which each element of $\mathbf{x}$ is referenced to time $t$ at the start of the sampling window. $M$ is the total number of satellites in view, $A^i$ and $\mathrm{s}^i$ are the $i$-th satellite amplitude and navigation data spread by the ranging code, respectively (Closas and Gusi-Amigó 2017; Peretic and Gao 2021). $f_{C/A}$ is the chipping rate of the spreading code of an L1 C/A signal. $n$ is the index for every element of $\mathbf{x}$, with $n = \left[0, \frac{1}{f_s}, \frac{2}{f_s}, \ldots, \Delta T - \frac{1}{f_s}\right]$, $f_s$ is the sampling frequency and $\Delta T$ is the sampling window or coherent integration time (Peretic 2019). $\phi_{\mathrm{code,t}}^i$ and $\varphi_{\mathrm{carr,t}}^i$ are the code phase in units of chips and carrier phase in radians of satellite $i$ at the current measurement sample time, $t$, respectively. $f_{\mathrm{L1}}$ is the L1 channel carrier frequency i.e., 1575.42 MHz, and $f_{\mathrm{code,t}}^i$ and $f_{\mathrm{carr,t}}^i$ are the code and carrier frequencies of satellite $i$ at time $t$, respectively. $noise$ represents the thermal noise of $\mathrm{x}(n)$ and is assumed to follow the Additive White Gaussian Noise (AWGN). On the other hand, $\mathbf{c}_i(\boldsymbol{\gamma}_j)$ is the vector of locally generated signal replica for satellite $i$ at the $j$-th candidate position, denoted by $\mathbf{c}_i(\boldsymbol{\gamma}_j) = \begin{bmatrix} \mathbf{c}_i[\boldsymbol{\gamma}_j, 0] \\ \vdots \\ \mathbf{c}_i\left[\boldsymbol{\gamma}_j, \Delta T - \frac{1}{f_s}\right] \end{bmatrix}$.

**Proposed DPE plug-in module**

This section will elaborate how the DPE plug-in module is integrated with the open-source GPS L1 C/A 2SP STL MATLAB SDR (SoftGNSS) (Borre et al. 2007). 2SP information, namely tracking code phase i.e., $\phi_{\mathrm{code,t}}^i$, signal transmission time, receiver local time i.e, $t$, satellite position from Least Squares, satellite clock bias, and Least Squares position solution, are used as input for the plug-in module. It might sound counterintuitive to use 2SP tracking estimates for DPE as DPE is supposed to be a single-step positioning algorithm that does not involve tracking. But the use of tracking measurements is merely for channel propagation (and ephemeris decoding in the 2SP part) in the intermediate frequency (IF) data. The possibility of using a tracking loop for propagating the IF channel has also been previously highlighted by Peretic (2019), who had developed his own open-source DPE SDR. A similar approach had also been previously introduced by Closas et al. (2015), who proposed using the delay locked loop (DLL) and phase lock loop (PLL) from 2SP tracking to synchronize the time and code phase. A flowchart detailing how the plug-in module is integrated into the SoftGNSS SDR is shown in Fig. 1.

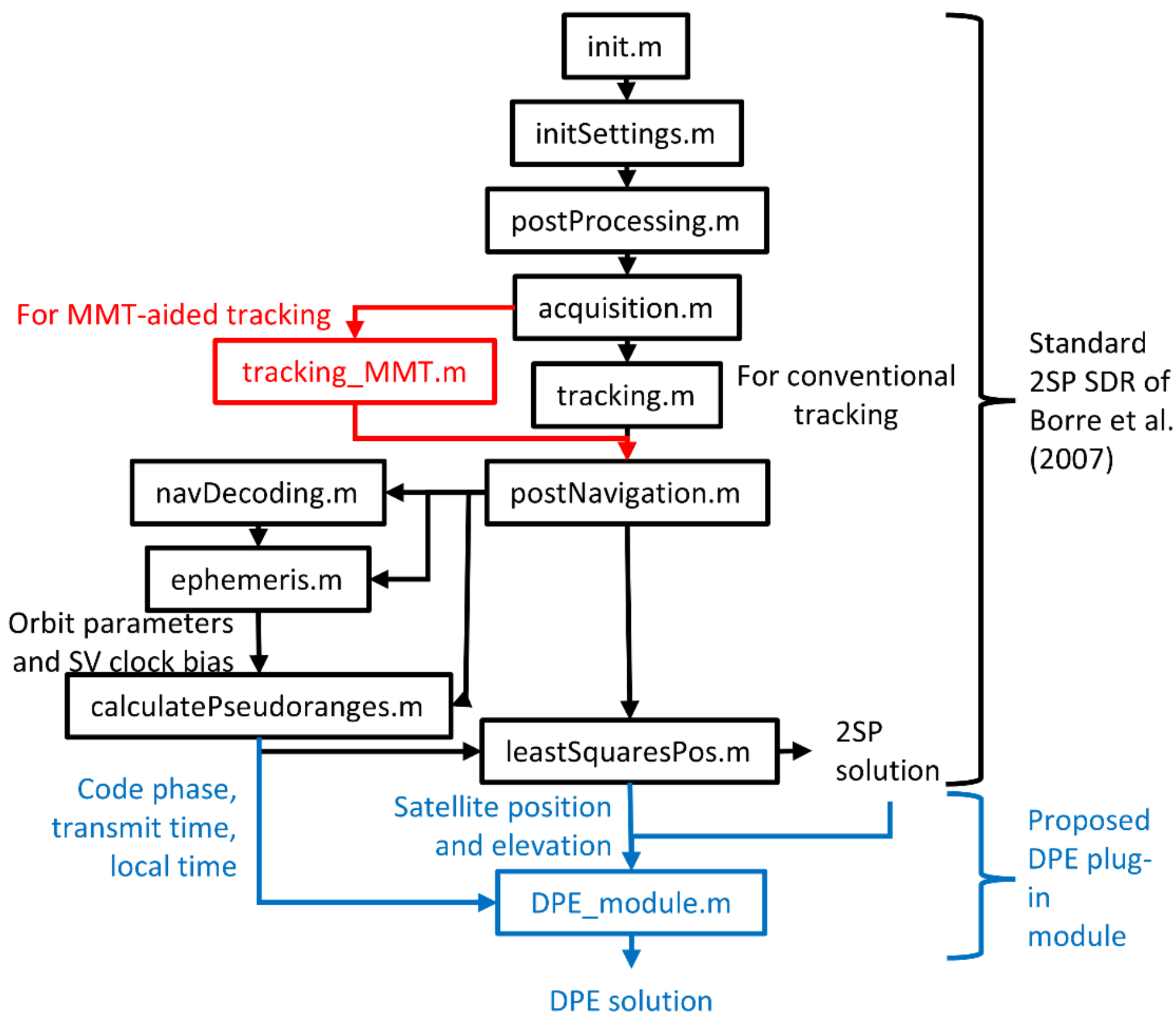

**Fig. 1** Flow chart of the proposed DPE plug-in module.

The proposed DPE module utilizes tracking measurements to generate the locally generated signal replicas for each candidate PVT, $\mathbf{c}_i(\boldsymbol{\gamma}_j)$. More specifically, each element of $\mathbf{c}_i$ will be presented as

$$c_i(\boldsymbol{\gamma}_j, n) = s^i\left\{\left(f_{C/A} + f^i_{\text{code,t}}\right)n + \phi^i_{\text{code,t}} + \Delta\phi^i_{\boldsymbol{\gamma}_j}\right\}\exp\left\{j2\pi\left(f_{\text{L1}} + f^i_{\text{carr,t}}\right)n + \varphi^i_{\text{carr,t}}\right\} \quad (4)$$

where $\Delta\phi^i_{\boldsymbol{\gamma}_j}$ is the difference in code phase between a specific candidate position with the estimated code phase at time $t$, which is computed with the following equation (Peretic and Gao 2021):

$$\Delta\phi^i_{\boldsymbol{\gamma}_j} = -f_{C/A}\left(t - t^i_{\text{transmit}}\right) + \frac{f_{C/A}}{c}\left(\left\|\mathbf{p}^{\mathrm{i}} - \mathbf{p}_j\right\| + \left(\delta t_j - \delta t^i\right)\cdot c + I + T\right) \quad (5)$$

where $t^i_{\text{transmit}}$ is the transmission time of satellite $i$. $\mathbf{p}^{\mathrm{i}}$ and $\mathbf{p}_j$ are the $i$-th satellite and $j$-th candidate position ECEF coordinates, respectively. $\delta t^i$ and $\delta t_j$ are the $i$-th satellite clock bias as well as the $j$-th candidate position clock bias, respectively. $I$ and $T$ are the ionospheric and tropospheric errors in units of meters, respectively, and $c$ is the speed of light (Peretic 2019). As seen in (1) - (5), the proposed DPE is currently configured to only estimate the receiver position and clock bias. As noted by Peretic and Gao (2021), estimation of position and velocity parameters can be decoupled. Thus, estimation of velocity parameters is not considered in this paper.

For (4), $\phi^i_{\text{code,t}}$ is obtained from *calculatePseudoranges*.m, which is the function for calculating pseudoranges, while $f^i_{\text{code,t}}$, $f^i_{\text{carr,t}}$ and $\varphi^i_{\text{carr,t}}$ are directly taken from the tracking function, *tracking*.m. Parameters for (5) i.e., $t^i_{\text{transmit}}$, $t$, $\mathbf{p}^{\mathrm{i}}$, $\delta t^i$, and $T$ are obtained from two-step positioning navigation estimates. Satellite elevation, to be used to correct the tropospheric error, as well as $\mathbf{p}^{\mathrm{i}}$, compensated for the Earth rotation are obtained from the function, *leastSquares*.m. $t^i_{\text{transmit}}$ and $t$, which are used to generate pseudoranges in 2SP, are obtained from *calculatePseudoranges*.m. $t$, which represents the local time in SoftGNSS, is kept uncompensated from Least Squares clock bias. Position as well as receiver clock bias estimate from *leastSquares*.m are also used as initialization for the grid of candidate position. Since the SoftGNSS codes were not configured to output the aforementioned parameters originally, the codes need to be slightly modified to output the required parameters for the DPE plug-in module.

The grid of candidate PVTs are by default configured to have 1 meter spacing between them. The grid spacing for the latitude-longitude-height estimation can be changed from *initSettings*.m under the parameter *settings.candPVT_spacing*. Yet, the spacing between the clock bias candidates is fixed at 1 meter. The PVT search space is set to have a span of ±30 meters for the latitude and longitude search space, ±50 meters for the height search space, and ±20 meters for clock bias by default. The candidate PVTs are centered around 2SP PVT estimate for every epoch to avoid DPE estimates from diverging in cases of strong interference or MP or NLOS effect.

The search space span can be edited in *initSettings*.m under the parameters *settings.DPE_latlong_span*, *settings.DPE_height_span*, and *settings.DPE_clkBias_span* (the parameter names are self-explanatory). The non-coherent and coherent integration time for DPE can also be changed through *initSettings*.m under the parameter *settings.DPE_nonCohInt* and *settings.DPE_CohInt* respectively, to improve the performance of DPE as the satellite correlations will be more filtered. The SoftGNSS v3.0 SDR has been updated to allow *settings.DPE_CohInt* to change the coherent integration time for both 2SP and DPE. For an apple-to-apple comparison between DPE and 2SP, a coherent integration time of 20 ms and 1 non-coherent integration for both DPE and 2SP are used throughout the results from datasets of GPS L1 C/A. The DPE-integrated SoftGNSS SDR requires ground truth coordinates in geodetic coordinates to output the positioning errors from both DPE and 2SP. Ground truth coordinates are labelled as *settings.gt_llh* at *initSettings*.m.

The proposed DPE can also be set to output the correlograms, plotted at the estimated DPE height, for each of the satellite correlations as well as the resulting non-coherent sum over the navigation domain (parameter *settings.DPE_plotCorrelogram*). Samples of the correlograms are shown in Fig. 2 below.

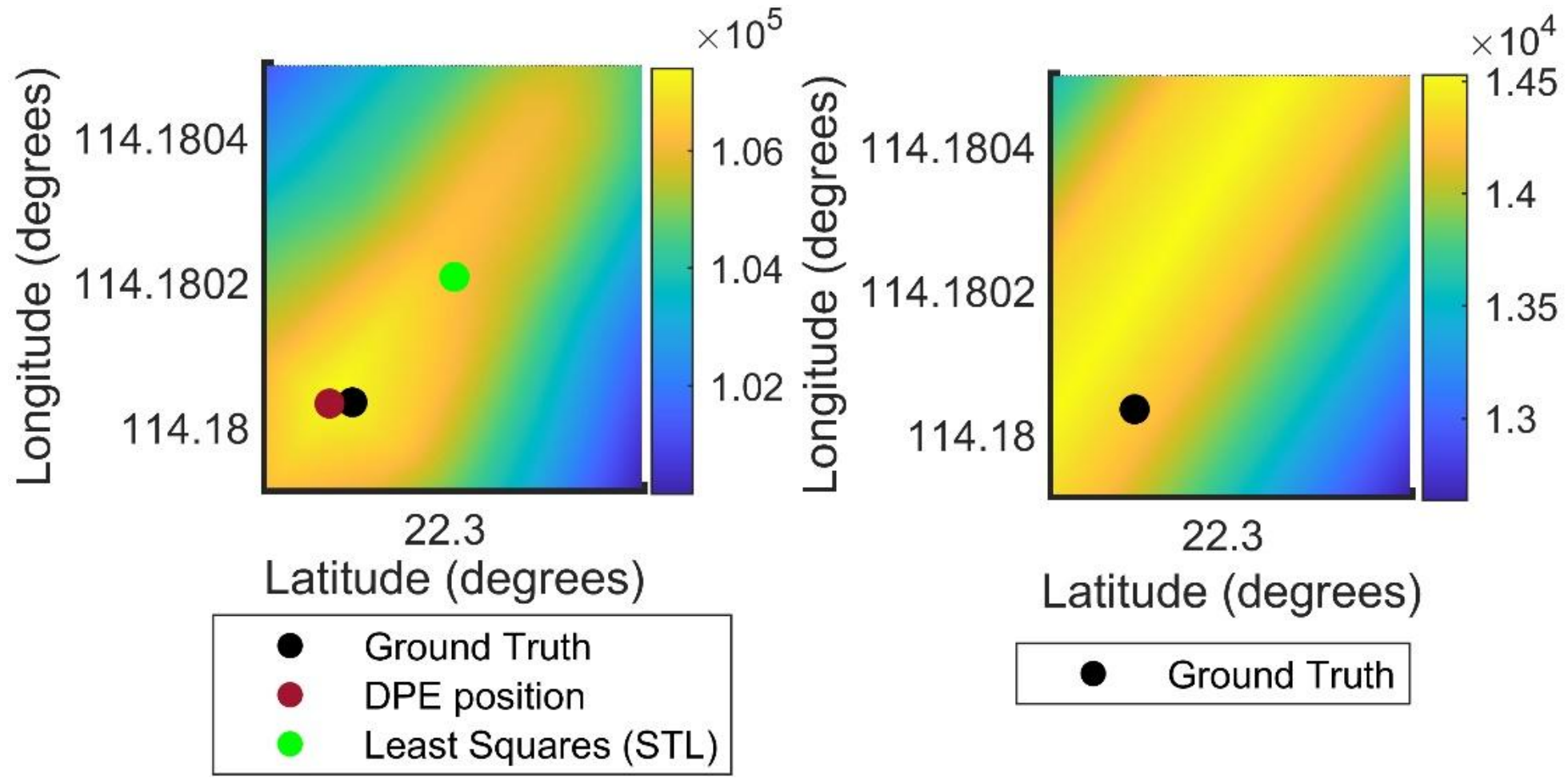

**Fig. 2** The final correlogram after summing up individual satellite correlations (Left) and the correlogram of a single satellite (Right), plotted across degrees latitude and longitude at estimated DPE height.

Instead of iteratively computing the signal correlations per every candidate position, the DPE plug-in module pre-calculates the correlations per every pre-determined chip spacing or code phase. This implementation was previously used by previous research on collective detection, which is another name for DPE, to save computational time (Axelrad et al. 2009, 2011; Cheong et al. 2011; Li et al. 2013). Since the pre-calculated correlations correspond to a specific code phase, the correlations for every candidate position are later given based on its code phase. Since it is virtually impossible to pre-calculate the correlations that fit perfectly to every candidate position code phase, the correlations will be interpolated. A loss in accuracy is expected from the use of interpolation. The pre-calculated correlations are done at spacings of chips/sample, computed with $f_{\mathrm{C/A}}/f_s$ by default for efficiency of the computations. For flexibility, the spacings between the pre-calculated correlations can be changed through *initSettings.m* under the parameter *settings.chipspacing_dpe_precalc*.

### Multipath Mitigation with DPE

Before exploring into the proposed MMT-DPE, the basics of how the proposed DPE plug-in module architecture can outperform 2SP against MP will first be explained. As mentioned above, the implementation of DPE will first pre-calculate the correlations per every code phase, with the code phase referenced to that produced from tracking. An illustration of the process is provided in the figure below.

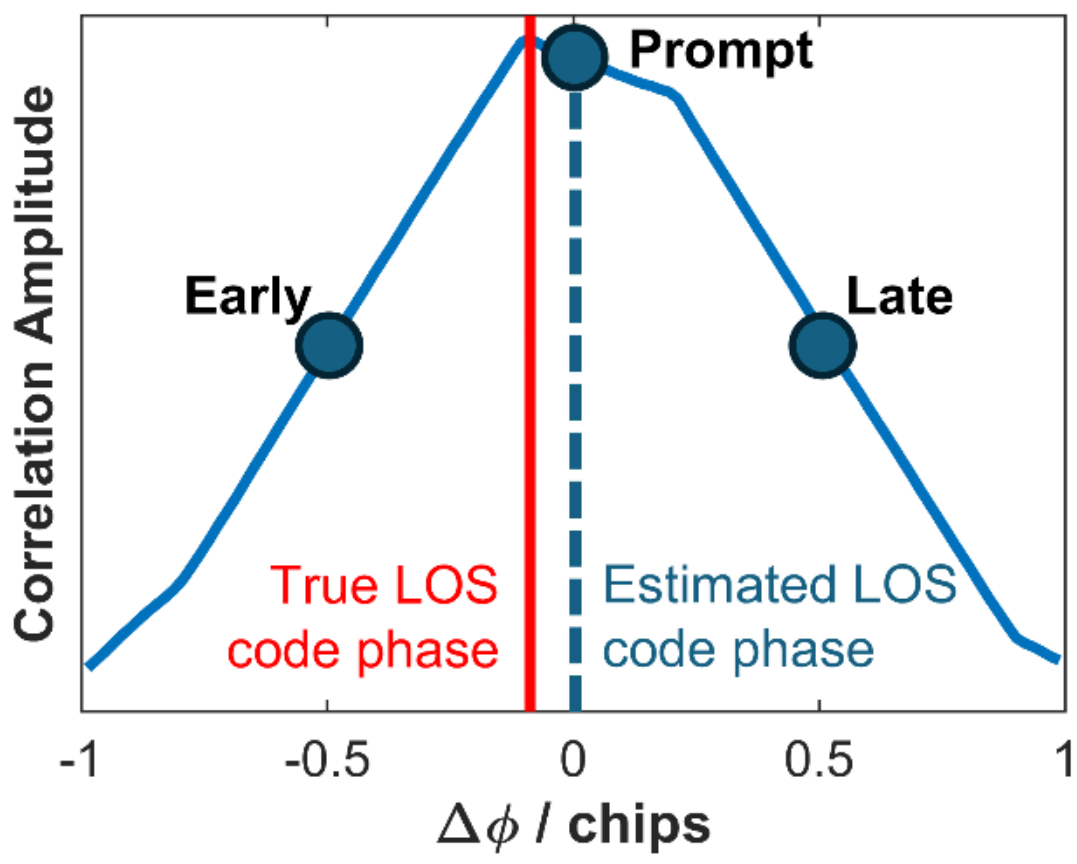

**Fig. 3** Relationship between $\Delta\phi$ and correlation value of a constructive MP-affected ACF in 2SP tracking. In this case, the relative amplitude of the reflected signal is less than the LOS path.

It is well known that MP deforms the ACF, which introduces errors in tracking (Xu et al. 2019b). Since the tracking loops aim to equalize the Early-Late correlator (E-L) values in tracking, it fails to track the LOS signal code phase correctly due to the ACF deformation from MP. Notice that the LOS peak, denoted by the red line in Fig. 3 is not aligned with the prompt correlator, which denotes the estimated LOS code phase by tracking i.e., zero mark of $\Delta\phi$. This difference is the pseudorange error from MP, and since 2SP uses this MP-affected pseudorange in its PVT computation, the error is propagated into the PVT estimation. On the contrary, DPE is *in theory* immune to the effects of both constructive and destructive MP if the relative signal amplitude of the reflected signal is lower than the LOS (Bialer et al. 2013).

Under the assumption that the reflected path is weaker than the LOS, the peak of the ACF still corresponds to the LOS peak i.e., the peak of the ACF still corresponds to the correct LOS code delay or pseudorange value. Thus, the highest correlation value is still given to candidate PVTs with the true range. Empirically, this also applies to destructive MP cases. But for a constructive MP, notice from Fig. 3 that the right-hand part of the ACF is elevated due to the constructive interference of the reflected signal with the LOS. With the effects of noise, the ACF peak may not correspond to the LOS peak in real cases. Xie and Petovello (2015) had also previously discovered that the ACF peak does not necessarily correspond to the LOS peak in real urban environments i.e., a stronger reflected signal than the LOS. In this case, DPE signal model will be misspecified as the assumption for the LOS peak being the highest in (1) becomes invalid, leading to poor performance from DPE (Bialer et al. 2013). A worst-case scenario of the ACF deformation in real urban environments is shown in Fig. 4 below.

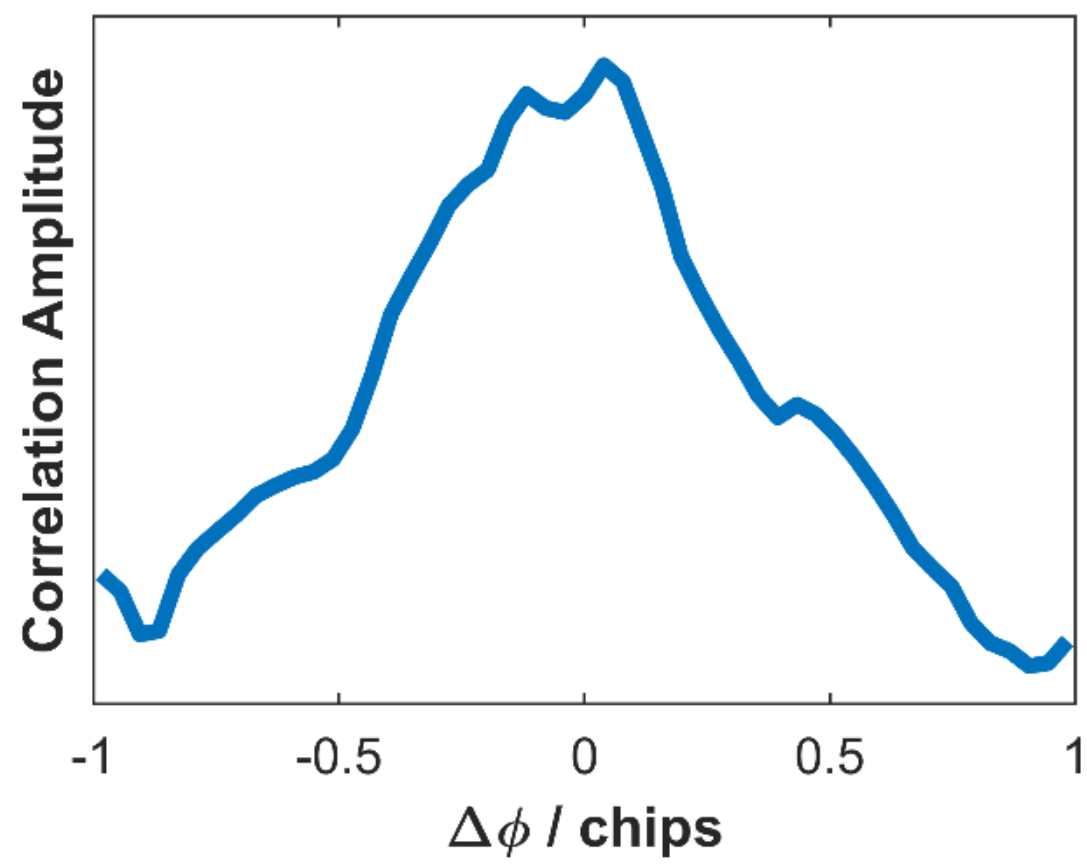

**Fig. 4** An example of an ACF from real urban data.

In addition, even simulated cases of MP have been shown to induce errors of tens of meters to DPE (Tang et al. 2024). Indeed, DPE remains robust against 2SP, but since wide range of research has investigated ways to reduce MP errors for 2SP, the prevalence of DPE as a robust positioning algorithm has been strongly diminished. As a result, an MMT-DPE is proposed.

## Multipath Mitigation Technology

Weill (2002) introduced MMT as a more computationally efficient alternative to his previously introduced MP estimating algorithm, the Minimum Mean-Square Error (MMSE), which required a six-fold integration to produce the delay estimates of the LOS path. Assuming a single-path MP reception, for a Doppler-compensated and navigation data free signal, the received baseband signal in equation (3), sampled at time *t*, can be simplified as the following

$$\mathrm{x}(t) = A^{\mathrm{LOS}} \cdot \mathrm{m}^{\mathrm{LOS}}\{t-\tau^{\mathrm{LOS}}\} \cdot e^{j\varphi_{\mathrm{carr,t}}^{\mathrm{LOS}}} + A^{\mathrm{NLOS}} \cdot \mathrm{m}^{\mathrm{NLOS}}\{t-\tau^{\mathrm{NLOS}}\} \cdot e^{j\varphi_{\mathrm{carr,t}}^{\mathrm{NLOS}}} + n(t) \quad (6)$$

where $\mathrm{m}\{\cdot\}$ represents the ranging code. $\tau^{\mathrm{LOS}}$ and $\tau^{\mathrm{NLOS}}$ represent the code delay of the LOS and reflected or NLOS path respectively (Weill 2002). $\varphi_{\mathrm{carr,t}}^{\mathrm{LOS}}$ and $\varphi_{\mathrm{carr,t}}^{\mathrm{NLOS}}$ represents the carrier phase of the LOS and reflected or NLOS path respectively. $A^{\mathrm{LOS}}$ and $A^{\mathrm{NLOS}}$ represents the amplitude of the LOS and reflected or NLOS path respectively. Separating the in-phase (Real) and quadrature (Imaginary) components yields

$$\mathrm{x}_{\mathrm{Real}}(t) = A^{\mathrm{LOS}} \cdot \mathrm{m}^{\mathrm{LOS}}\{t-\tau^{\mathrm{LOS}}\} \cdot \cos\left(\varphi_{\mathrm{carr,t}}^{\mathrm{LOS}}\right) + A^{\mathrm{NLOS}} \cdot \mathrm{m}^{\mathrm{NLOS}}\{t-\tau^{\mathrm{NLOS}}\} \cdot \cos\left(\varphi_{\mathrm{carr,t}}^{\mathrm{NLOS}}\right) + n_{\mathrm{Real}}(t) \quad (7)$$

$$\mathrm{x}_{\mathrm{Imaginary}}(t) = A^{\mathrm{LOS}} \cdot \mathrm{m}^{\mathrm{LOS}}\{t-\tau^{\mathrm{LOS}}\} \cdot \sin\left(\varphi_{\mathrm{carr,t}}^{\mathrm{LOS}}\right) + A^{\mathrm{NLOS}} \cdot \mathrm{m}^{\mathrm{NLOS}}\{t-\tau^{\mathrm{NLOS}}\} \cdot \sin\left(\varphi_{\mathrm{carr,t}}^{\mathrm{NLOS}}\right) + n_{\mathrm{Imaginary}}(t) \quad (8)$$

where $n_{\mathrm{Real}}(t)$ and $n_{\mathrm{Imaginary}}(t)$ represent noise in the real and imaginary components respectively, both independent, real-valued, and follows AWGN. To estimate the MP parameters, MMT involves the minimization of the following cost function (Weill 2002).

$$\Gamma = \begin{array}{l} \int_0^{\mathrm{T}} \begin{bmatrix} \mathrm{x}_{\mathrm{Real}}(t) - A^{\mathrm{LOS}} \cdot \mathrm{m}^{\mathrm{LOS}}\{t - \tau^{\mathrm{LOS}}\} \cdot \cos(\varphi_{\mathrm{carr,t}}^{\mathrm{LOS}}) \\ -A^{\mathrm{NLOS}} \cdot \mathrm{m}^{\mathrm{NLOS}}\{t - \tau^{\mathrm{NLOS}}\} \cdot \cos(\varphi_{\mathrm{carr,t}}^{\mathrm{NLOS}}) \end{bmatrix}^2 dt \\ + \int_0^{\mathrm{T}} \begin{bmatrix} \mathrm{x}_{\mathrm{Imaginary}}(t) - A^{\mathrm{LOS}} \cdot \mathrm{m}^{\mathrm{LOS}}\{t - \tau^{\mathrm{LOS}}\} \cdot \sin(\varphi_{\mathrm{carr,t}}^{\mathrm{LOS}}) \\ - A^{\mathrm{NLOS}} \cdot \mathrm{m}^{\mathrm{NLOS}}\{t - \tau^{\mathrm{NLOS}}\} \cdot \sin(\varphi_{\mathrm{carr,t}}^{\mathrm{NLOS}}) \end{bmatrix}^2 dt \end{array} \quad (9)$$

where T is the end of the sampling window. $\Gamma$ typically involves a non-linear optimization with six-parameter search space consisting of $\tau^{\mathrm{LOS}}$, $\tau^{\mathrm{NLOS}}$, $\varphi_{\mathrm{carr,t}}^{\mathrm{LOS}}$, $\varphi_{\mathrm{carr,t}}^{\mathrm{NLOS}}$, $A^{\mathrm{LOS}}$, and $A^{\mathrm{NLOS}}$. But instead, Weill (2002) used the following invertible transformation

$$\mathrm{A} = A^{\mathrm{LOS}} \cdot \cos(\varphi_{\mathrm{carr,t}}^{\mathrm{LOS}}) \quad \mathrm{C} = A^{\mathrm{LOS}} \cdot \sin(\varphi_{\mathrm{carr,t}}^{\mathrm{LOS}})$$

$$\mathrm{B} = A^{\mathrm{NLOS}} \cdot \cos(\varphi_{\mathrm{carr,t}}^{\mathrm{NLOS}}) \quad \mathrm{D} = A^{\mathrm{NLOS}} \cdot \sin(\varphi_{\mathrm{carr,t}}^{\mathrm{NLOS}}) \quad (10)$$

which transforms (9) into

$$\Gamma = \int_0^{\mathrm{T}} \left[\mathrm{x}_{\mathrm{Real}}{}^2(t) + \mathrm{x}_{\mathrm{Imaginary}}{}^2(t)\right] dt + (\mathrm{A}^2 + \mathrm{B}^2 + \mathrm{C}^2 + \mathrm{D}^2) \cdot R_{\mathrm{mm}}(0) - 2 \cdot \mathrm{A} \cdot R_{\mathrm{Real,m}}(\tau^{\mathrm{LOS}}) - 2 \cdot \mathrm{B} \cdot R_{\mathrm{Real,m}}(\tau^{\mathrm{LOS}}) + 2 \cdot \mathrm{A} \cdot \mathrm{B} \cdot R_{\mathrm{mm}}(\tau^{\mathrm{LOS}} - \tau^{\mathrm{NLOS}}) - 2 \cdot \mathrm{C} \cdot R_{\mathrm{Imag,m}}(\tau^{\mathrm{LOS}}) - 2 \cdot \mathrm{D} \cdot R_{\mathrm{Imag,m}}(\tau^{\mathrm{LOS}}) + 2 \cdot \mathrm{C} \cdot \mathrm{D} \cdot R_{\mathrm{mm}}(\tau^{\mathrm{LOS}} - \tau^{\mathrm{NLOS}}) \quad (11)$$

where $R_{\mathrm{mm}}$ is the autocorrelation function between two local replicas, $R_{\mathrm{Real,m}}$ and $R_{\mathrm{Imag,m}}$ are the cross-correlation between the local code replicas with (7) and (8) respectively (Chen et al. 2013). This ingenious transformation allows the original six-dimensional problem of (9) to turn into a two-dimensional optimization, through the minimization of (11) with respect to A, B, C, and D. Said minimization can be achieved by taking the partial derivatives of (11) with respect to said parameters

$$\frac{\partial \Gamma}{\partial \mathrm{A}} = 2 \cdot \mathrm{A} \cdot R_{\mathrm{mm}}(0) - R_{\mathrm{Real,m}}(\tau^{\mathrm{LOS}}) + 2 \cdot \mathrm{B} \cdot R_{\mathrm{mm}}(\tau^{\mathrm{LOS}} - \tau^{\mathrm{NLOS}}) = 0$$

$$\frac{\partial \Gamma}{\partial \mathrm{B}} = 2 \cdot \mathrm{B} \cdot R_{\mathrm{mm}}(0) - R_{\mathrm{Real,m}}(\tau^{\mathrm{NLOS}}) + 2 \cdot \mathrm{A} \cdot R_{\mathrm{mm}}(\tau^{\mathrm{LOS}} - \tau^{\mathrm{NLOS}}) = 0$$

$$\frac{\partial \Gamma}{\partial \mathrm{C}} = 2 \cdot \mathrm{C} \cdot R_{\mathrm{mm}}(0) - R_{\mathrm{Imag,m}}(\tau^{\mathrm{LOS}}) + 2 \cdot \mathrm{D} \cdot R_{\mathrm{mm}}(\tau^{\mathrm{LOS}} - \tau^{\mathrm{NLOS}}) = 0$$

$$\frac{\partial \Gamma}{\partial \mathrm{D}} = 2 \cdot \mathrm{D} \cdot R_{\mathrm{mm}}(0) - R_{\mathrm{Imag,m}}(\tau^{\mathrm{NLOS}}) + 2 \cdot \mathrm{C} \cdot R_{\mathrm{mm}}(\tau^{\mathrm{LOS}} - \tau^{\mathrm{NLOS}}) = 0 \quad (12)$$

For each potential (candidate) values of $\tau^{\text{LOS}}$ and $\tau^{\text{NLOS}}$, (12) allows the computation of their corresponding A, B, C, and D, which together with $\tau^{\text{LOS}}$ and $\tau^{\text{NLOS}}$ acts as inputs for (11). Our implementation of MMT uses a grid-based method, in which a grid of $\tau^{\text{LOS}}$ and $\tau^{\text{NLOS}}$ pairs are first generated and the pair of $\tau^{\text{LOS}}$ and $\tau^{\text{NLOS}}$ which minimizes (11) will be MMT's estimate of the LOS and NLOS delays. Since $\tau^{\text{NLOS}}$ is always larger than $\tau^{\text{LOS}}$, pair delays with $\tau^{\text{NLOS}}$smaller than the $\tau^{\text{LOS}}$ is not considered (Blanco-Delgado and Nunes 2012).

To improve the performance of MMT, the following constraint is used on the amplitudes, in the form of a Lagrange multiplier (Weill 2002).

$$\frac{A^{\text{NLOS}}}{A^{\text{LOS}}} \leq 0.8 \qquad (13)$$

To reduce the computational load, an iterative grid-search algorithm is used for MMT, in which the $\tau^{\text{LOS}}$ and $\tau^{\text{NLOS}}$ search space and grid spacing are iteratively reduced. This method is inspired by the multi-resolution collective detection method by Li et al. (2013), in which the candidate PVT grids are sequentially reduced from coarse to fine. A band of multicorrelators are used to initialize the code delay grids. The code delay peak of the multicorrelators will then be used to initialize the $\tau^{\text{LOS}}$ and $\tau^{\text{NLOS}}$ grids. This is intended to not compromise the dynamic performance of an MMT tracking, and at the same time, allow the MMT to initialize at the best estimate of $\tau^{\text{LOS}}$.

From there, MMT will initially run on grid spacings of 0.1 chips, and the chip spacings will be iteratively reduced by half, with a final resolution of 0.00625 chips. This will indeed generate a maximum bias of 0.003125 chips, or around 0.916 meters for MMT estimates, which in the framework of single-point positioning, can be considered acceptable. Fig. 5 elaborates the workflow of MMT.

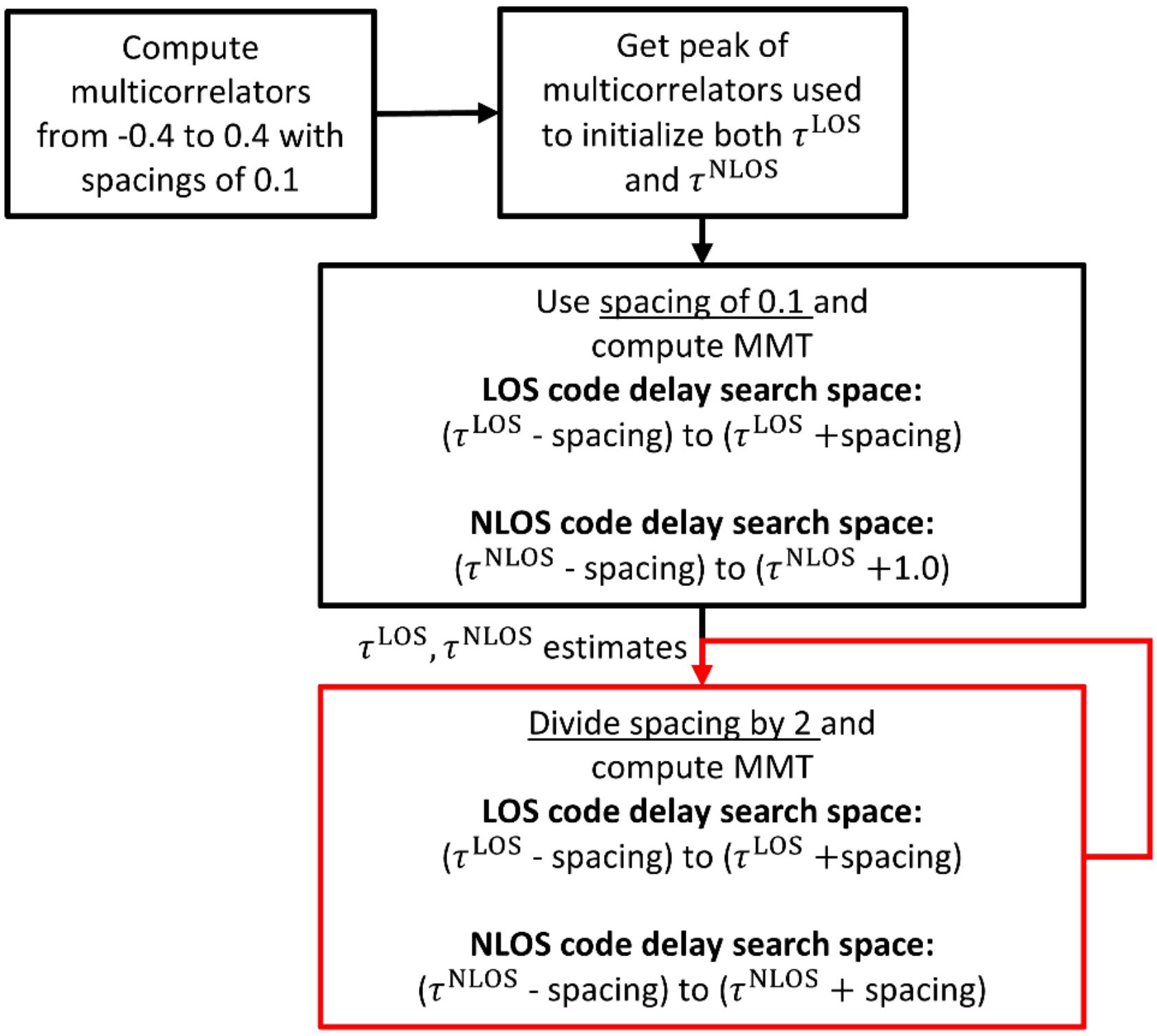

**Fig. 5** Workflow of the implemented MMT algorithm. The iteration to reduce the grid

resolution (marked in red) is repeated 4 times, resulting in a final resolution of 0.00625 chips.

### Proposed MMT-DPE

The proposed MMT-DPE methodology aims to solve DPE misspecified signal model due to MP by virtually replacing DPE cost function with that of MMT i.e., (9) or (11). In other words, DPE cost function in (1) would be revised into the following for MMT-DPE.

$$\hat{\boldsymbol{\gamma}} = \arg \min_{\boldsymbol{\gamma},\boldsymbol{\tau}^{\mathrm{NLOS}},\mathbf{a}^{\mathrm{LOS}},\mathbf{a}^{\mathrm{NLOS}}} \sum_{i=1}^{\mathrm{M}} \left\| \mathbf{x} - \mathbf{c}_i(\boldsymbol{\gamma})\mathrm{a}^{\mathrm{LOS},i} - \mathbf{c}_i\left(\tau_i^{\mathrm{NLOS}}\right)\mathrm{a}^{\mathrm{NLOS},i} \right\|^2 \quad (14)$$

or equivalently

$$\hat{\boldsymbol{\gamma}} = \arg \max_{\boldsymbol{\gamma},\boldsymbol{\tau}^{\mathrm{NLOS}},\mathbf{a}^{\mathrm{LOS}},\mathbf{a}^{\mathrm{NLOS}}} \Big\{ \sum_{i=1}^{\mathrm{M}} \mathbf{x}^{\mathrm{H}}\mathbf{c}_i(\boldsymbol{\gamma})\mathrm{a}^{\mathrm{LOS},i} + \mathbf{x}^{\mathrm{H}}\mathbf{c}_i\left(\tau_i^{\mathrm{NLOS}}\right)\mathrm{a}^{\mathrm{NLOS},i} + \left(\mathrm{a}^{\mathrm{LOS},i}\right)^{\mathrm{H}}\left(\mathbf{c}_i(\boldsymbol{\gamma})\right)^{\mathrm{H}}\mathbf{x} - \left(\mathrm{a}^{\mathrm{LOS},i}\right)^{\mathrm{H}}\left(\mathbf{c}_i(\boldsymbol{\gamma})\right)^{\mathrm{H}}\mathbf{c}_i(\boldsymbol{\gamma})\mathrm{a}^{\mathrm{LOS},i} - \left(\mathrm{a}^{\mathrm{LOS},i}\right)^{\mathrm{H}}\left(\mathbf{c}_i(\boldsymbol{\gamma})\right)^{\mathrm{H}}\mathbf{c}_i\left(\tau_i^{\mathrm{NLOS}}\right)\mathrm{a}^{\mathrm{NLOS},i} + \left(\mathrm{a}^{\mathrm{NLOS},i}\right)^{\mathrm{H}}\left(\mathbf{c}_i\left(\tau_i^{\mathrm{NLOS}}\right)\right)^{\mathrm{H}}\mathbf{x} - \left(\mathrm{a}^{\mathrm{NLOS},i}\right)^{\mathrm{H}}\left(\mathbf{c}_i\left(\tau_i^{\mathrm{NLOS}}\right)\right)^{\mathrm{H}}\mathbf{c}_i(\boldsymbol{\gamma})\mathrm{a}^{\mathrm{LOS},i} - \left(\mathrm{a}^{\mathrm{NLOS},i}\right)^{\mathrm{H}}\left(\mathbf{c}_i\left(\tau_i^{\mathrm{NLOS}}\right)\right)^{\mathrm{H}}\mathbf{c}_i\left(\tau_i^{\mathrm{NLOS}}\right)\mathrm{a}^{\mathrm{NLOS},i} \Big\} \quad (15)$$

where $\mathrm{a}^{\mathrm{LOS},i}$ and $\mathrm{a}^{\mathrm{NLOS},i}$ are the complex amplitudes of the LOS and reflected paths for satellite $i$, respectively, while $\mathbf{c}_i(\tau_i^{\mathrm{NLOS}})$ is the vector of locally generated signal replica for the reflected path for satellite $i$.

Since the proposed DPE is a closed-loop implementation i.e., using tracking measurements to propagate the channel, the channel propagation must also be compensated for MP for effective MMT implementation for DPE. Thus, MMT is implemented in 2SP tracking. The discriminator output in 2SP tracking, in which the SoftGNSS is configured to use the normalized Early-minus-Late power discriminator, will directly be replaced with $\tau^{\mathrm{LOS}}$ estimates from MMT. This would inherently produce more accurate pseudorange measurements for 2SP, which would automatically yield MMT-2SP.

(15) also highlights the additional estimation of the complex amplitudes of the LOS and reflected path, as well as the reflected path code delay. But practically, only estimation of the PVT and reflected path code delay is needed since computation of (15) can be used the same way as MMT cost function in (11) i.e., the complex amplitudes can be obtained through partial derivatives of the cost function with respect to the complex amplitudes for a specific LOS and reflected path code delay pair as elaborated in (12).

Since MMT is implemented in tracking, $\mathrm{a}^{\mathrm{LOS,i}}$,$\mathrm{a}^{\mathrm{NLOS},i}$, as well as $\tau_i^{\mathrm{NLOS}}$ can already be obtained from tracking. As the SoftGNSS v3.0 is a post-processing algorithm which processes tracking of the whole data first before positioning, MMT-DPE computational overhead can be reduced by skipping the computation of the partial derivatives and directly using the $\mathrm{a}^{\mathrm{LOS},i}$,$\mathrm{a}^{\mathrm{NLOS},i}$, as well as $\tau_i^{\mathrm{NLOS}}$ from tracking to compute (15) for the pre-calculation of the correlations for MMT-DPE positioning.

## Data Collection Method

Evaluation of the proposed DPE module will use both real and simulated static GNSS data. Three front-ends were involved in collection of real data; National Instruments Universal Software Radio Peripheral (NI USRP) B210, NI USRP N210, and NSL Stereo.

For the NI USRP front-ends, a ZYACF-L004 antenna was used, powered by a bias-tee. The NI USRP was connected to an Ubuntu laptop with GNU Radio to record and save the down-converted IF data. For the NI USRP B210, a sampling frequency of 20 MHz was used, with bandwidth of 2 MHz. Conversely, the NI USRP N210 used a wider 10 MHz bandwidth with a sampling frequency of 10 MHz. For the N210, since it does not draw power from the laptop like the B210, a battery pack was required to power it. This battery pack was also used to power the bias-tee. On the other hand, datasets collected with the NSL Stereo used an Allystar AGR6303 antenna. The NSL Stereo has a 26 MHz sampling frequency and an 8 MHz bandwidth (Xu et al. 2019a). For the simulated urban datasets, a Spirent simulator was used to generate datasets with known MP and NLOS parameters. The simulator was connected to a LabSat 3W front-end, configured with 58 MHz of sampling frequency and 56 MHz of front-end bandwidth. Ground truth for all the collected urban datasets were taken with a u-Blox F9P receiver, which uses a multi-constellation Real-Time Kinematic (RTK) positioning to obtain centimeter-level accuracy.

## Results and Discussion

The results section will first investigate the performance of pure DPE and 2SP. 2SP scalar tracking is configured to use 0.6 E-L chip spacing for all of the results. In cases with no MP or NLOS, both 2SP and DPE are expected to have comparable positioning performance (Closas and Gusi-Amigó 2017). Real open-sky BeiDou B1I data was collected from the roof of R-core building of The Hong Kong Polytechnic University to illustrate the case. Fig. 6 shows the skymask as well as the positioning result from both DPE and 2SP. Since the BeiDou result is of open-sky data, and merely used as proof of concept that the DPE module can be integrated with SDRs for other GNSS constellations, the coherent integration time for both DPE and 2SP are kept at SoftGNSS v3.0 default i.e., 1 ms for the following BeiDou result.

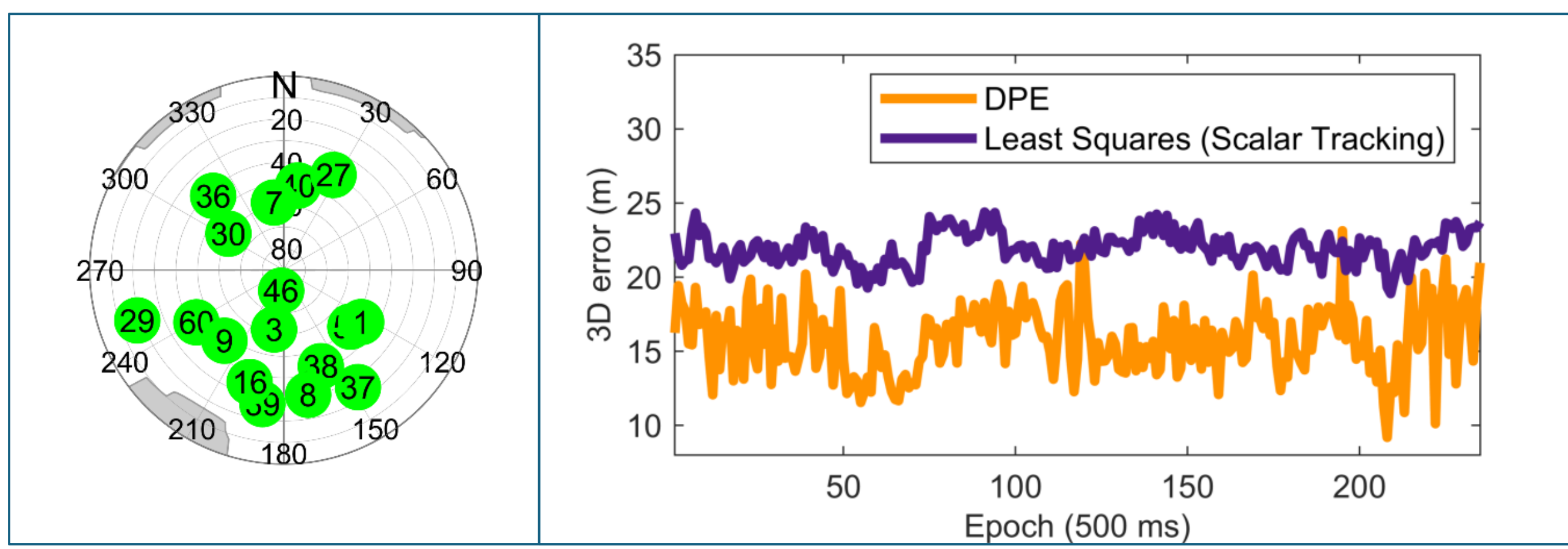

**Fig. 6** Positioning results from DPE and 2SP with BeiDou B1I open-sky dataset, with an average 3D mean error of 21.94 meters and 15.81 meters from 2SP and DPE, respectively. This dataset was collected with the NI USRP N210.

Even though collected in an open-sky environment, the fluctuation in the positioning error from Fig. 6 indicates the existence of MP, making it more of a light urban environment instead of open-sky. The result from Fig. 6 clearly favors DPE, due to its inherent MP mitigation property. Similar results can be observed with a GPS L1 C/A open-sky dataset, but due to the much lesser number of satellites in view, the positioning improvement from DPE is much less compared to that seen in BeiDou. Fig. 7 shows the open-sky results with GPS L1 C/A.

Evaluating DPE under controlled environments, a Spirent simulator was used to generate two simulated urban conditions; medium urban and harsh urban. In the simulated light urban environment, there are a total of 8 GPS L1 C/A satellites, with one simulated constructive MP satellite of 0.1 chip relative delay (and 0.5 relative amplitude) for the reflected signal and one NLOS satellite with 0.2 chip additional delay. Fig. 8 presents the positioning result for the simulated light urban dataset.

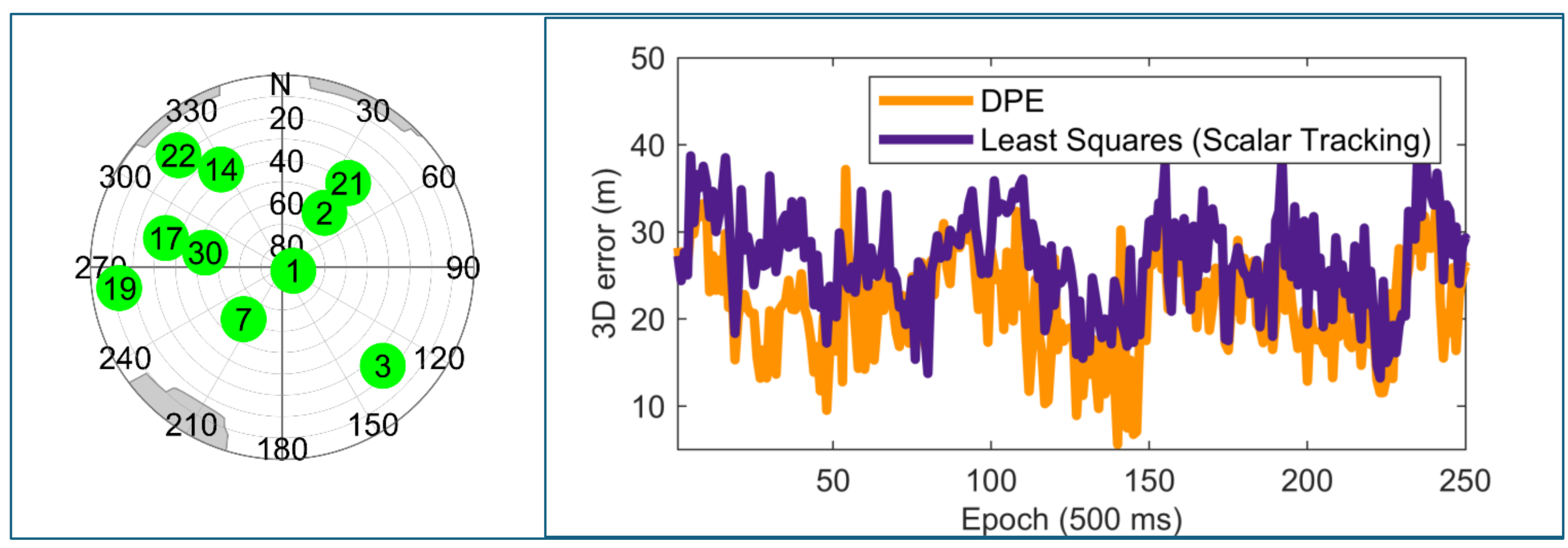

**Fig. 7** Positioning results from DPE and 2SP with GPS L1 C/A open-sky dataset, with an average 3D mean error of 26.71 meters and 21.49 meters from 2SP and DPE, respectively. This dataset was collected with the NI USRP N210.

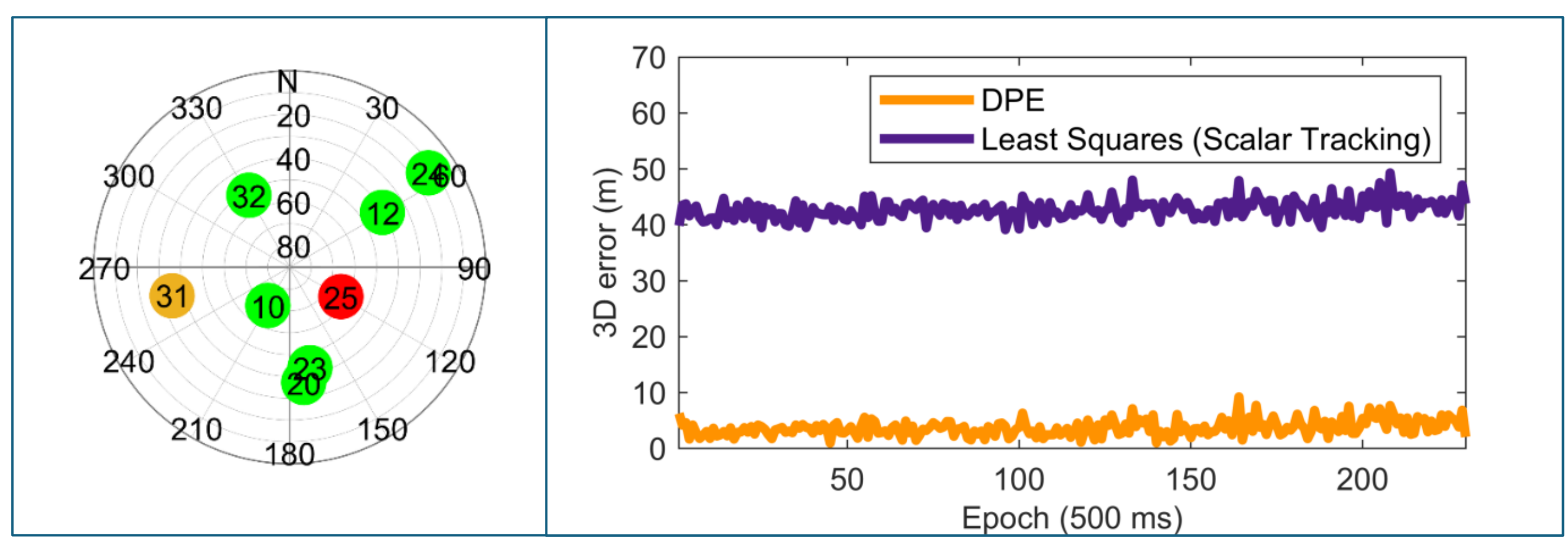

**Fig. 8** Positioning results from DPE and 2SP under a simulated GPS L1 C/A medium urban condition, with an average 3D mean error of 42.98 meters and 3.68 meters from 2SP and DPE, respectively. PRN 25 (highlighted in red) is the simulated NLOS satellite, while PRN 31 (highlighted in orange) is the simulated MP.

On the contrary, there are a total of 9 GPS L1 C/A satellites for the simulated harsh urban dataset. Three satellites are constructive MP, with PRN12 configured to have 0.05 chip relative delay (and 0.8 relative amplitude), PRN 26 with 0.1 chip relative delay (and 0.5 relative amplitude), and PRN23 with 0.3 chip relative delay (and 0.4 relative amplitude) for the reflected signal. A single NLOS satellite was also simulated, configured with 0.3 chip of additional delay. The results are elaborated in Fig. 9. Results from the Spirent simulated datasets clearly show DPE positioning robustness in comparison to 2SP, with 91.38% mean outperformance with the simulated medium urban dataset. But for the harsh urban, the geometry of the MP and NLOS satellites allows for the pseudorange errors to cancel out, leading to similar performance from 2SP and DPE.

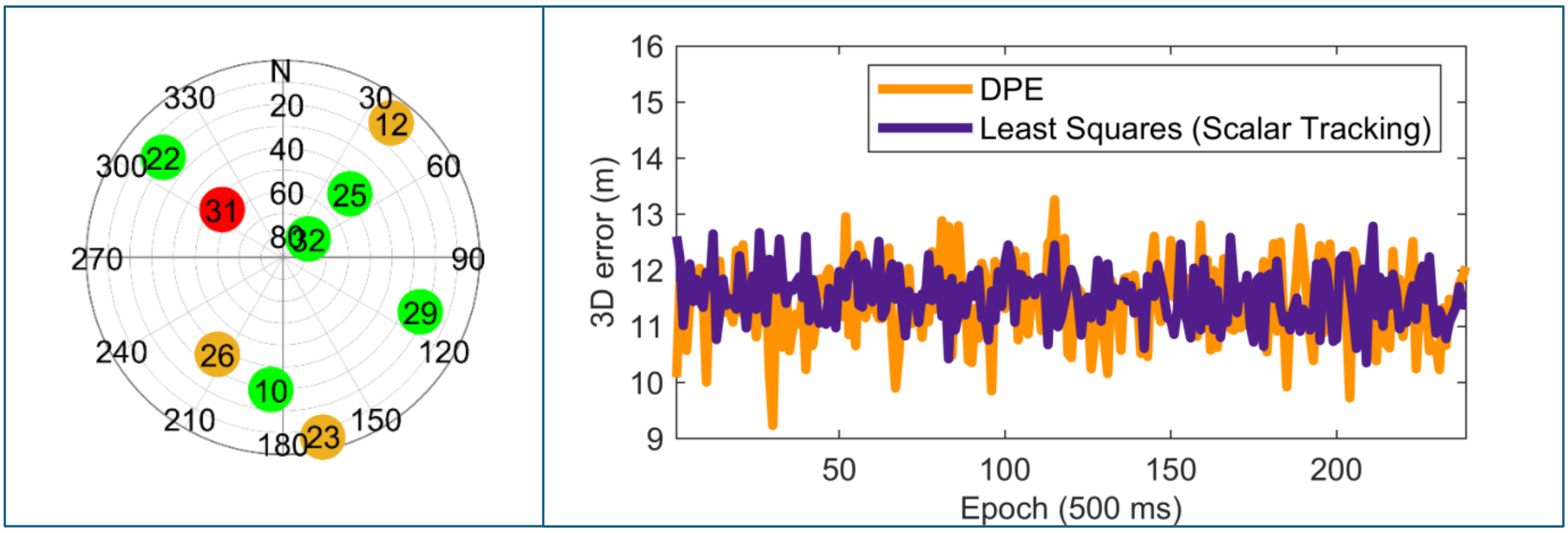

**Fig. 9** Positioning results from DPE and 2SP under a simulated GPS L1 C/A harsh urban condition, with an average 3D mean error of 11.55 meters and 11.41 meters from 2SP and DPE, respectively. PRN 31 (highlighted in red) is the simulated NLOS satellite, while PRN 12, 23 and 26 (highlighted in orange) are the simulated MP satellites.

While previously, the resilience of GNSS DPE against MP has been thoroughly established, it is revealed that it has strong resilience against NLOS as well, supporting the arguments on Direct Position Determination (DPD) (Amar and Weiss 2005; Bialer et al. 2013). The findings from the simulated datasets are further supported by real GPS L1 C/A data with one NLOS satellite. Fig. 10 provides the result from a GPS L1 C/A medium urban dataset collected in East Tsim Sha Tsui, close to The Hong Kong Polytechnic University. With PRN 22 blocked by a nearby building (denoting NLOS reception), DPE manages to offer a 54.70% improvement in mean error compared to 2SP.

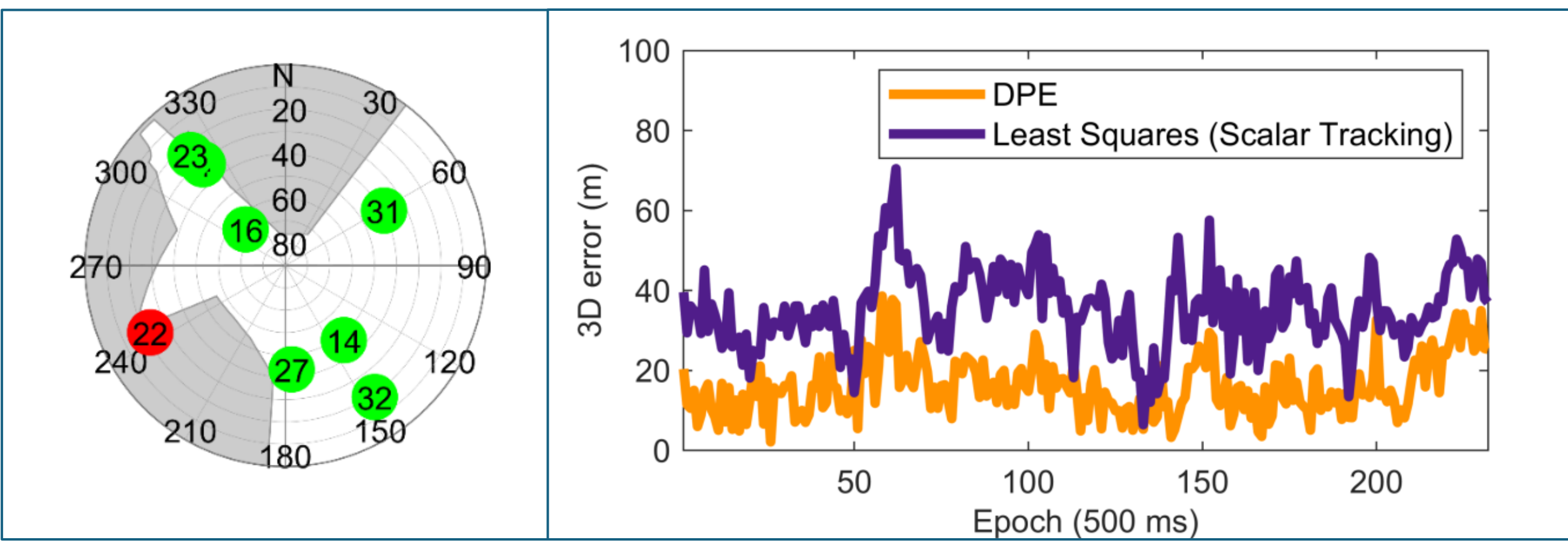

**Fig. 10** Positioning results from DPE and 2SP from a medium urban GPS L1 C/A dataset collected in East Tsim Sha Tsui, with an average 3D mean error of 35.30 meters and 15.99 meters from 2SP and DPE, respectively. PRN 22 (highlighted in red) is the NLOS satellite. This dataset was collected with NSL Stereo.

The research findings on DPE resilience against NLOS also concurs with the author's previous findings on an DPE-based positioning scheme, in which it solving the PVT in the navigation domain is crucial for NLOS resilience as it allows the exclusion of the error-affected satellite measurements from contributing to the final PVT (Vicenzo et al. 2024).

The case can be further illustrated by digging deeper into the correlograms result for an NLOS satellite. Fig. 11 shows the comparison between the correlograms of a high-elevated satellite and NLOS satellite from the medium urban dataset in Fig. 10. Fig. 11 clearly illustrates how the correlations of an NLOS satellite deviate from the ground truth. And in contrast, for a high-elevated satellite (which for the sake of argument, is assumed to be LOS), the correlations converge to the ground truth. On top of that, Fig. 11 clearly shows that the correlations of an NLOS satellite are much weaker compared to that of the highly elevated, which inherently gives a lower weighting to the NLOS satellite during the PVT estimation.

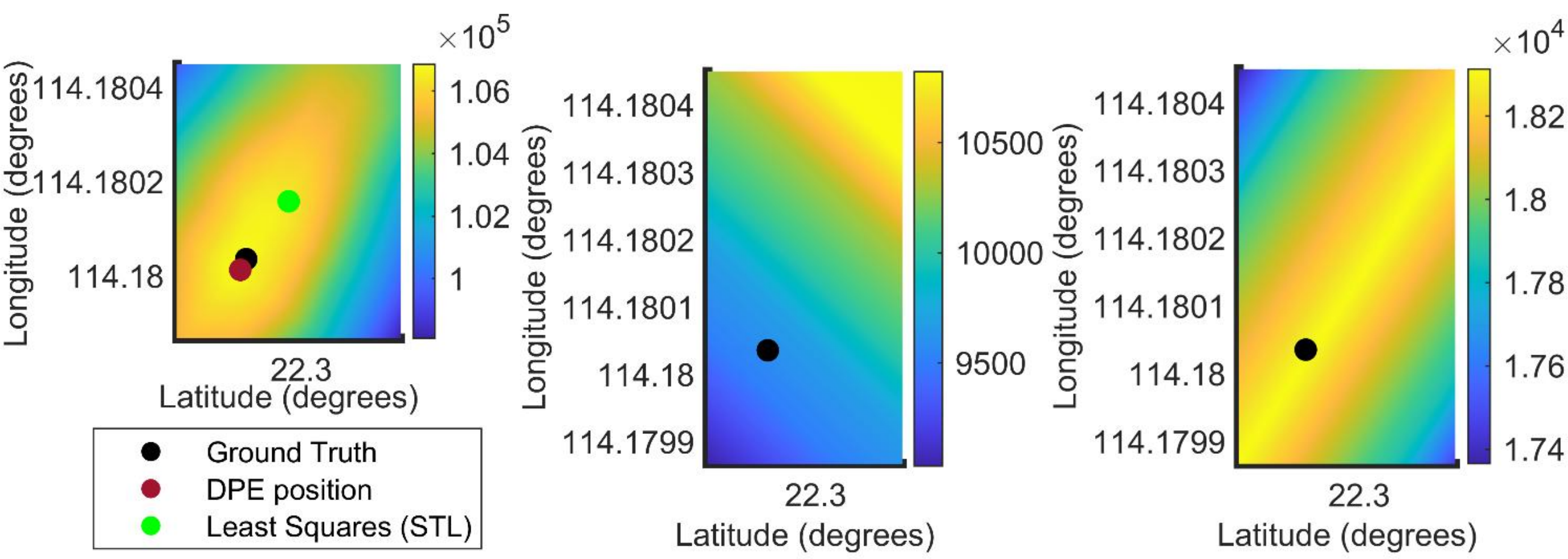

**Fig. 11** Correlogram of an NLOS satellite, PRN 22 (Center) in Fig. 10 and that of a highly elevated satellite, PRN 16 (Right) in Fig. 10. The final correlogram after summing up individual satellite correlations (Left) shows the position estimate from DPE remains close to the ground truth unlike 2SP.

Hence, NLOS satellite correlations do not contribute to the global maxima or in other words, its measurements are excluded from the final PVT estimation of DPE, and that final DPE PVT estimate made use of the correlations from the other satellites in view (Vicenzo et al. 2023; Amar and Weiss 2005; Bialer et al. 2013).

Tang et al. (2024) had presented that DPE manages to obtain much more accurate positioning even if the majority of signals received are MP. For the sake of completion, Fig. 12 illustrates the positioning error from the harsh urban Spirent data when only PRN 22, 12, 23, and 26 are left i.e., 3 MP signals and 1 LOS. DPE manages to consistently obtain a much more accurate position in comparison to 2SP, highlighting that the proposed DPE implementation meets the expected theoretical DPE MP mitigation capability previously elaborated above.

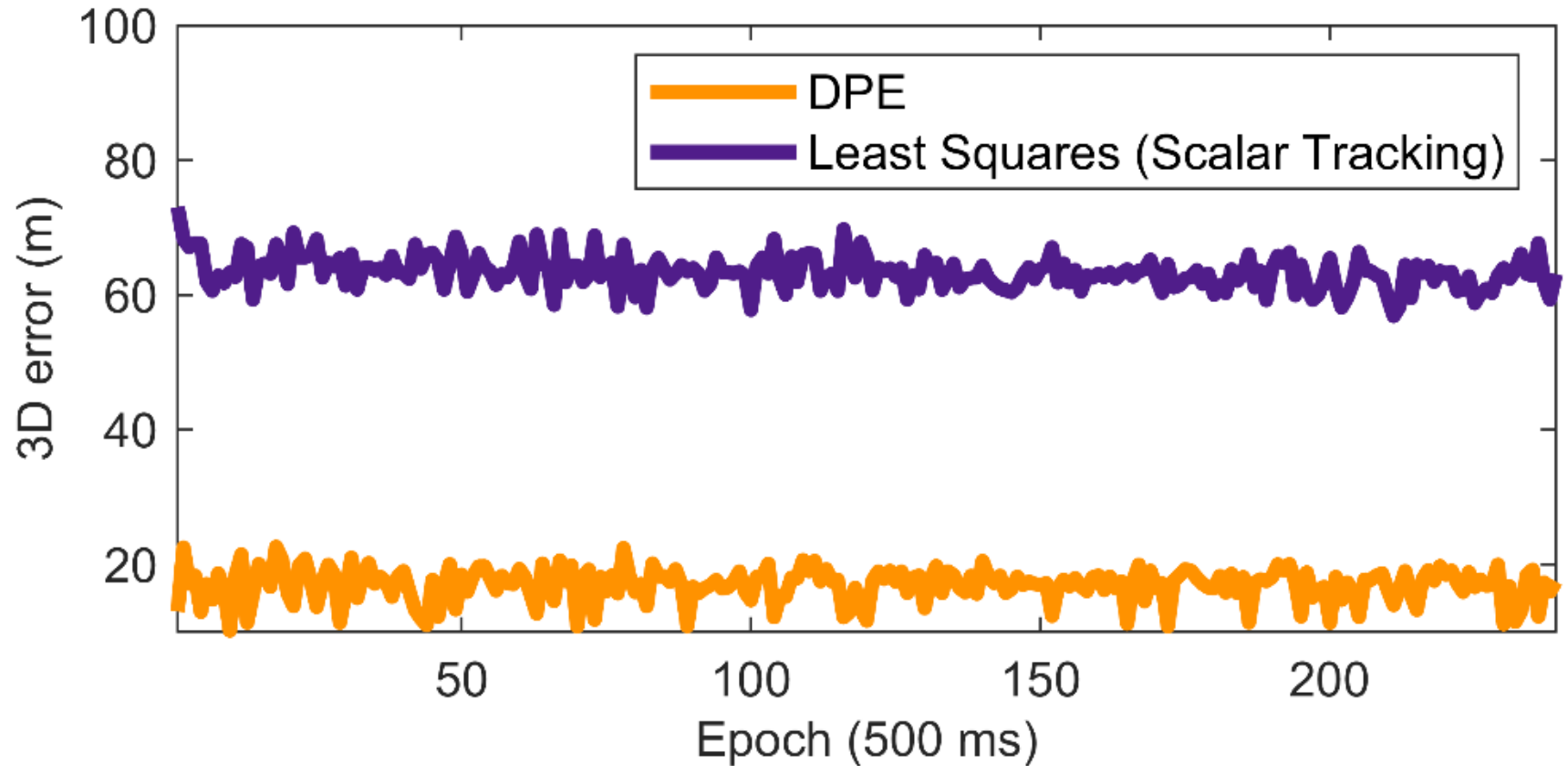

**Fig. 12** Positioning results from DPE and 2SP when there is only a single LOS satellite and 3 MP satellite i.e., an extremely harsh urban case, with an average 3D mean error of 63.35 meters and 16.94 meters from 2SP and DPE, respectively.

Moving into the results of the MMT-DPE, the performance of MMT estimates is analysed with the medium urban data from East Tsim Sha Tsui shown previously in Fig. 10. Since the $\tau^{\mathrm{LOS}}$ estimates from MMT are used to replace the normalized Early-minus-Late power discriminator output in tracking, the discriminator output from three satellites is compared to the $\tau^{\mathrm{LOS}}$ from MMT in Fig. 13. Fig. 13 clearly shows that $\tau^{\mathrm{LOS}}$ from MMT has a much lower variance in 2SP tracking error compared to the discriminator.

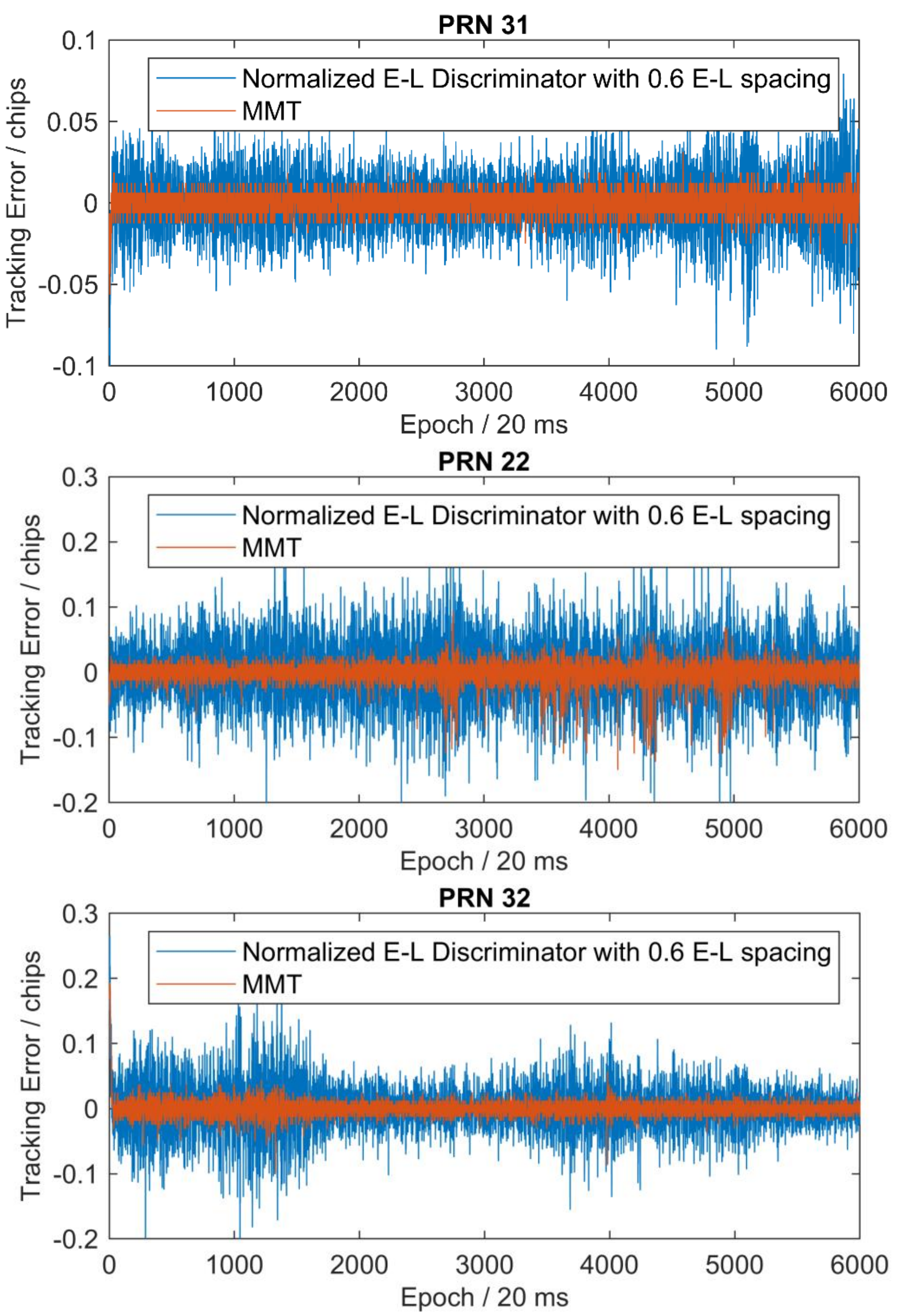

**Fig. 13** Comparison of normalized Early-minus-Late power discriminator output with the $\tau^{\mathrm{LOS}}$ from MMT when integrated into 2SP tracking for real medium urban data

The effect of using MMT will directly be investigated under the real medium urban data from East Tsim Sha Tsui in Fig. 10 with the NLOS satellite (PRN22) excluded. The positioning result is elaborated on Fig. 14 below. MMT-DPE 3D error is recorded to be 5.85 meters and MMT-2SP being 5.84 meters. This presents not only a huge improvement for 2SP, which uncompensated by MMT, has a mean 3D error of 21.32 meters, but also for DPE, which has a mean error of 9.03 meters without MMT.

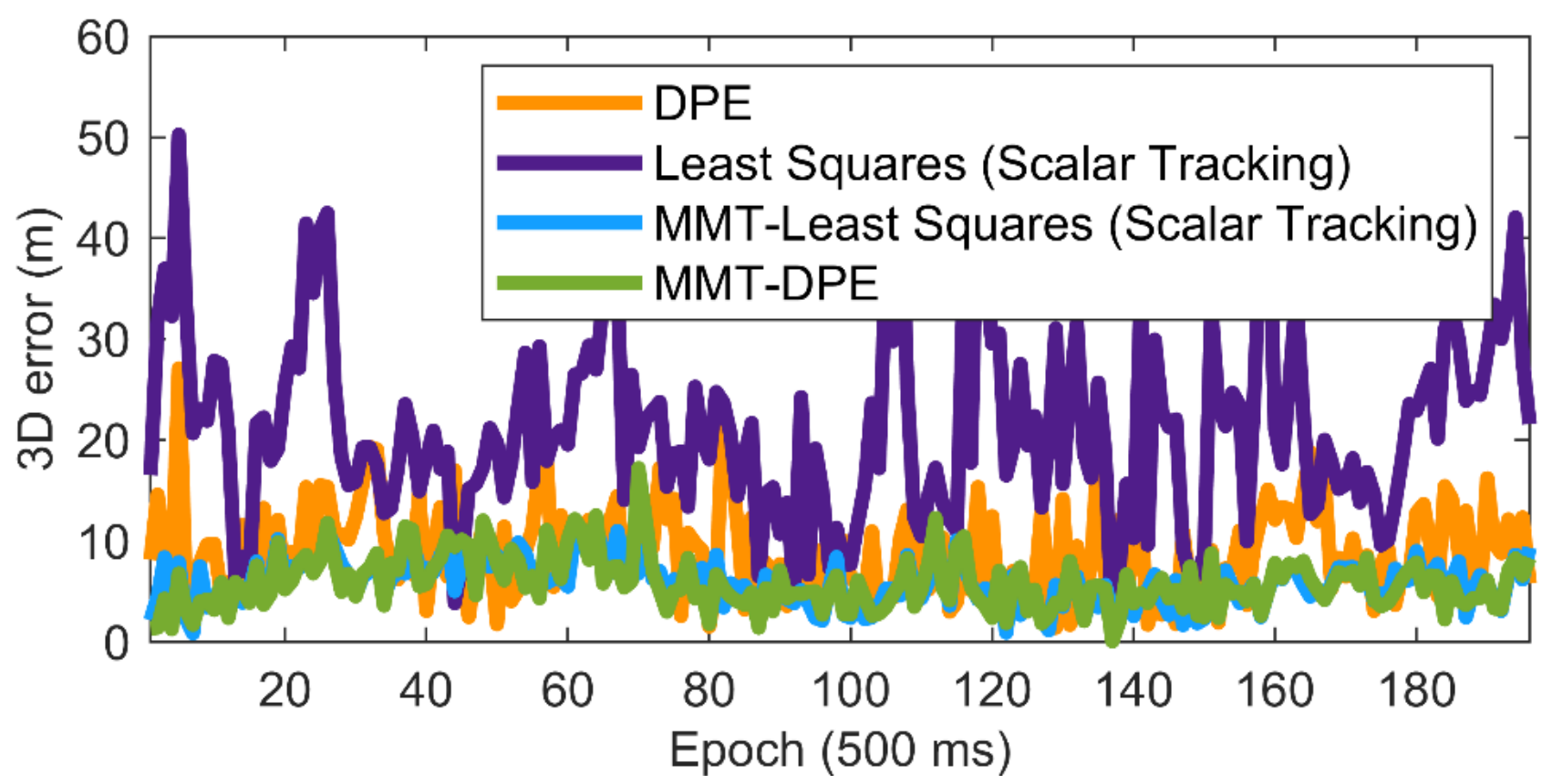

**Fig. 14** Comparison of positioning results from MMT-DPE and 2SP with pure DPE and 2SP with real medium urban data from East Tsim Sha Tsui with NLOS satellite excluded.

It must be reminded that even though both 2SP and DPE now perform similarly now that the MP reception is taken into account into the signal models, NLOS reception is also prominent in urban environments, which cannot be compensated with MMT for both DPE and 2SP. But as illustrated above, the inherent superiority of DPE to NLOS allows the MMT-DPE positioning to remain accurate when NLOS is introduced.

To illustrate this case, the positioning results of the real medium urban data from Fig. 10, which has a single NLOS satellite, is shown in Fig. 15. The original DPE outperformance of 54.70% against pure 2SP is now increased to 78.09%, with MMT-DPE has a mean 3D error of 6.71 meters while 2SP is 30.64 meters, thus proving that the MMT-DPE is the more suitable option for urban positioning in comparison to pure DPE, pure 2SP, and MMT-2SP.

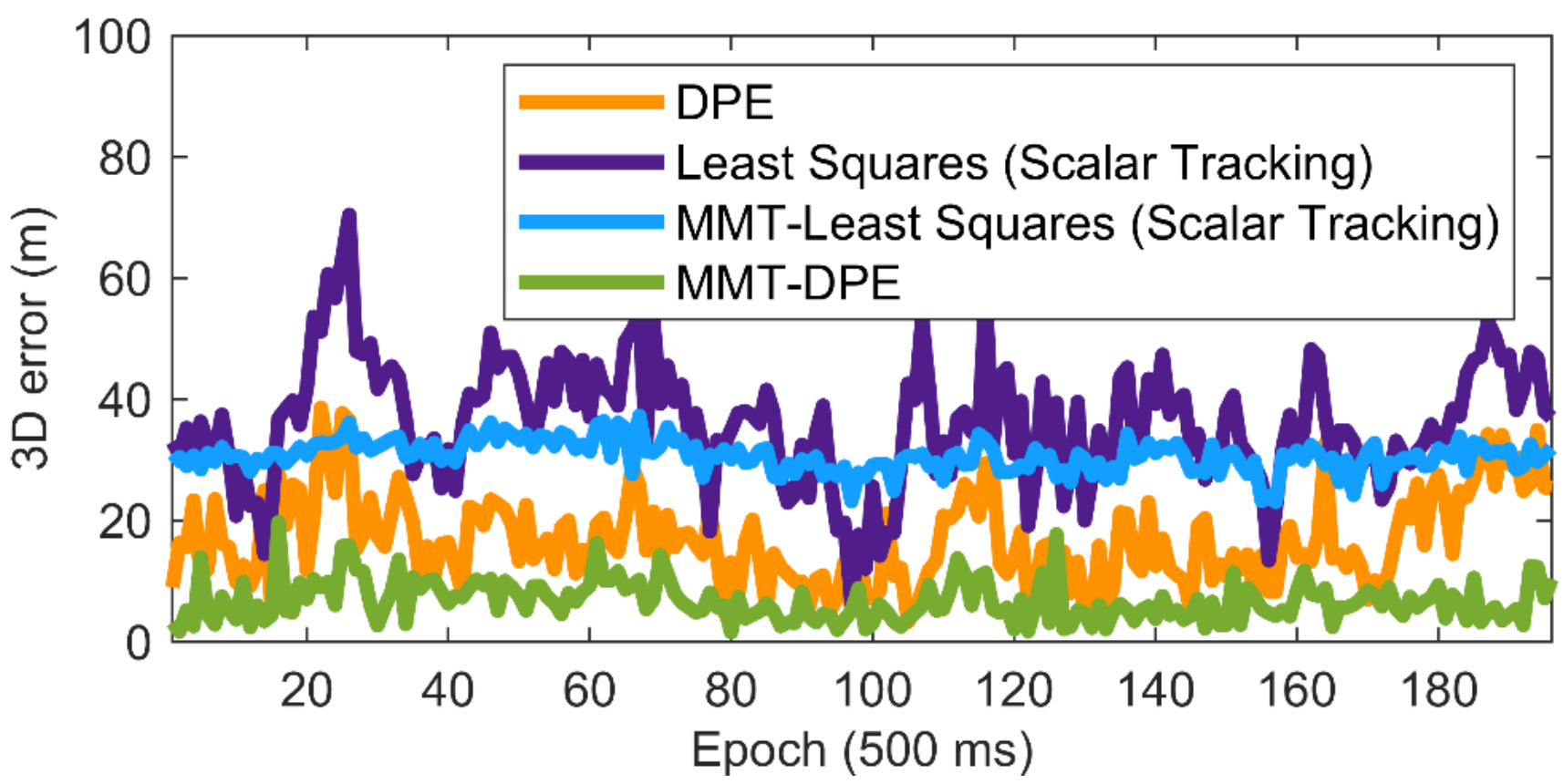

**Fig. 15** Positioning results from DPE and 2SP from a medium urban GPS L1 C/A dataset collected in East Tsim Sha Tsui, with a single NLOS satellite.

## Conclusion

We have presented a novel implementation of a DPE receiver, which has been made open-sourced to the GNSS community through the GitHub page https://github.com/Sergio-Vicenzo/GPSL1-DPEmodule. With the aim of better familiarization and understanding of DPE, the proposed DPE is programmed in the user-friendly and easy-to-learn programming language, MATLAB and as a module that can be integrated into existing MATLAB 2SP SDRs. Currently, the DPE module is made open source integrated into the SoftGNSS MATLAB GPS L1 C/A 2SP SDR from Borre et al. (2007), which is based on STL. But integration of the proposed DPE module is not restricted to an STL-based receiver and can be integrated with other 2SP architectures, such as VTL. Additionally, the GNSS constellation is also not restricted to GPS L1 C/A, but is also integrable with SDRs working with other BPSK-modulated GNSS signals. The results section has shown results from using BeiDou B1I signal, which has a similar signal structure with GPS L1 C/A.

Results clearly prove that the proposed DPE plug-in module still harnesses the established advantages of DPE and consistently outperforms 2SP in the tested real and simulated datasets. DPE manages to achieve 91.38% positioning improvement in a simulated medium urban and 54.70% for real medium urban. Even when three out of four satellites are MP i.e., an extremely harsh urban case, DPE manages to maintain an average error below 20 meters whereas 2SP error goes beyond 60 meters.

A variant of DPE, dubbed MMT-DPE has also been introduced to improve DPE resilience against MP, which has also been made available at the same GitHub page. Though DPE manages to consistently achieve lower positioning errors compared to 2SP, recent research has shown that under severe MP conditions, its error can still explode to tens of meters like 2SP does. As a result, MMT-DPE was introduced to better suit DPE for applications in urban environments.

Under MP-only conditions, both MMT-DPE and MMT-2SP perform similarly. But since DPE also has a natural NLOS mitigation property, MMT-DPE consistently outperforms MMT-2SP when both MP and NLOS exist. Under real urban data with both MP and NLOS, MMT-DPE manages to maintain roughly the same error as without NLOS, which further highlights its NLOS mitigation property. In contrast, an MMT-2SP seems to not significantly improve 2SP positioning, due to the existence of NLOS. As a result, MMT-DPE manages to outperform both MMT-2SP and unaided 2SP by up to 78.09%, making it the preferable choice for urban positioning.

## Declarations

Ethics approval and consent to participate: Not applicable.

Consent for publication: Not applicable.

Availability of data and materials: The proposed open-source DPE plug-in module as well as its MMT derivative is made available by accessing https://github.com/Sergio-Vicenzo/GPSL1-DPEmodule

Competing interests: The authors declare that they have no competing interests.

Funding: This study was supported by the National Natural Science Foundation of China (NSFC) under Grant 62103346.

Author contributions: Sergio Vicenzo and Bing Xu developed the idea and methodology. Sergio Vicenzo developed the software. Sergio Vicenzo and Bing Xu conducted formal analysis and investigation. Sergio Vicenzo wrote the main manuscript, conducted the experiment, and prepared the figures and tables. Bing Xu conducted the review, editing and supervision of the work. All authors read and approved the final manuscript.

Acknowledgements: While not directly involved in the research, the authors would like to thank Prof. Dennis M. Akos of the University of Colorado Boulder for his input on having the same coherent and non-coherent integration times for an apple-to-apple comparison between 2SP and DPE during the author's previous presentation in ION GNSS+ 2023. We would also like to thank Dr.-Ing. Jürgen Dampf of Universität der Bundeswehr München, as well as Dr. Pau Closas of the Northeastern University for their valuable suggestions on our DPE implementation at ION GNSS+ 2023.

## Author Biographies

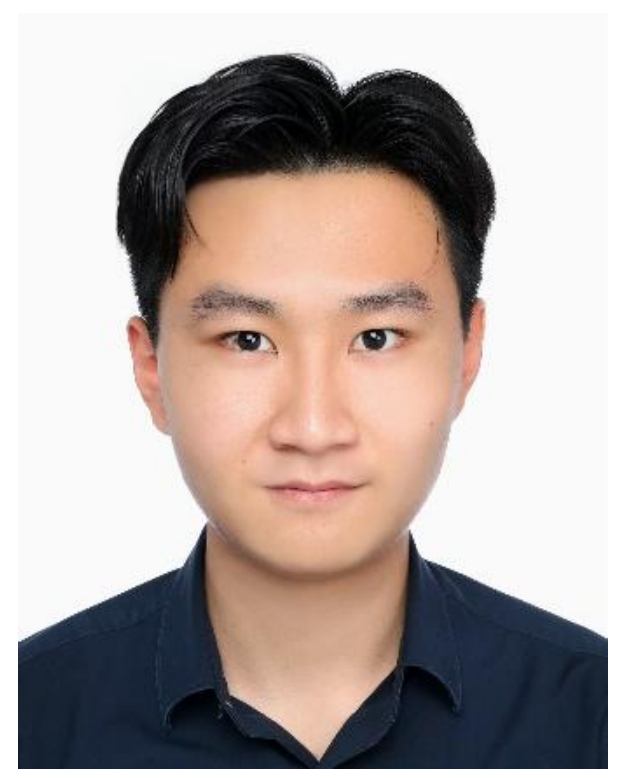

**Sergio Vicenzo** is currently a Ph.D. candidate at the Department of Aeronautical and Aviation Engineering, The Hong Kong Polytechnic University. He received first-class honors in Bachelor of Engineering in Aviation Engineering from the same university in 2022. His research interests include GNSS urban navigation and positioning with direct position estimation.

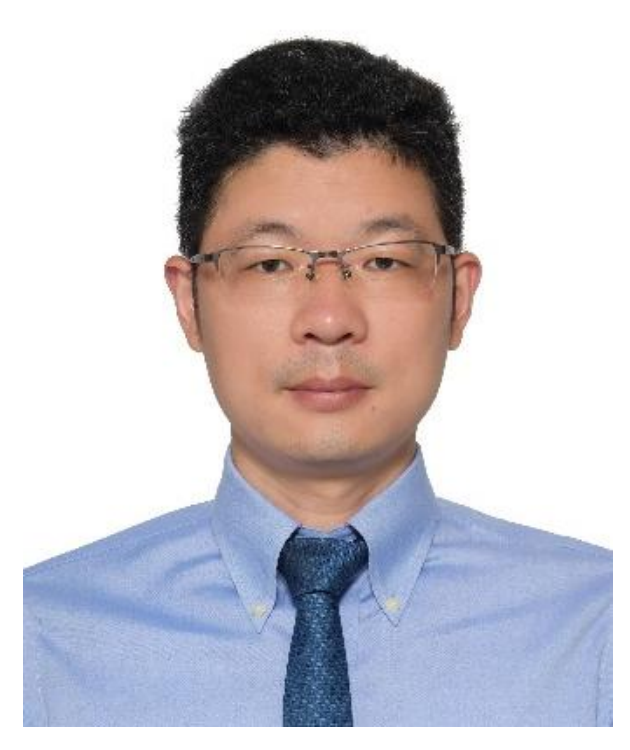

**Bing Xu** received the BEng and Ph.D. degrees in network engineering and control science and engineering from the Nanjing University of Science and Technology, Nanjing, China, in 2012 and 2018, respectively. He is currently an Assistant Professor with the Department of Aeronautical and Aviation Engineering, The Hong Kong Polytechnic University. His current research interests include GNSS signal processing in software-defined global navigation satellite system receivers.